\documentclass{jfm}
\usepackage{natbib,amsmath,amssymb,graphicx}
\usepackage[usenames,dvipsnames,svgnames,table]{xcolor}

\ifCUPmtlplainloaded \else
  \checkfont{eurm10}
  \iffontfound
    \IfFileExists{upmath.sty}
      {\typeout{^^JFound AMS Euler Roman fonts on the system,
                   using the 'upmath' package.^^J}%
       \usepackage{upmath}}
      {\typeout{^^JFound AMS Euler Roman fonts on the system, but you
                   dont seem to have the}%
       \typeout{'upmath' package installed. JFM.cls can take advantage
                 of these fonts,^^Jif you use 'upmath' package.^^J}%
      }
  \else
  \fi
\fi

\ifCUPmtlplainloaded \else
  \checkfont{msam10}
  \iffontfound
    \IfFileExists{amssymb.sty}
      {\typeout{^^JFound AMS Symbol fonts on the system, using the
                'amssymb' package.^^J}%
       \usepackage{amssymb}%
       \let\le=\leqslant  
       \let\ge=\geqslant  
      }{}
  \fi
\fi

\ifCUPmtlplainloaded \else
  \IfFileExists{amsbsy.sty}
    {\typeout{^^JFound the 'amsbsy' package on the system, using it.^^J}%
     \usepackage{amsbsy}}
    {\providecommand\boldsymbol[1]{\mbox{\boldmath $##1$}}}
\fi

\usepackage{overpic}
\usepackage{amsmath,mathrsfs}
\usepackage{psfrag}
\usepackage{amsmath,amssymb,amsfonts,bbm}
\usepackage[bbgreekl]{mathbbol}
\usepackage{mathrsfs}

\newcommand{\tens}[1]{\boldsymbol{\mathsf{#1}}}
\renewcommand{\vec}[1]{\ensuremath{\mbox{\boldmath$#1$}}}
\newcommand{\ma}[1]{\ensuremath{\mathbb{#1}}}

\begin{document}

\title{Time-dependent lift and drag on a rigid body in a viscous steady linear flow} 
\author[F. Candelier, B. Mehlig and J. Magnaudet]
{Fabien Candelier$^1$, Bernhard Mehlig$^2$ and Jacques Magnaudet$^3$}
\affiliation{$^1$Aix-Marseille Univ., CNRS, IUSTI  (Institut Universitaire des Syst\`emes Thermiques et Industriels)
F-13013 Marseille, France\\
$^2$ Department of Physics, Gothenburg University,
 SE-41296 Gothenburg, Sweden\\
$^3$ Institut de M\'ecanique des Fluides de Toulouse (IMFT), Universit\'e de Toulouse, CNRS, INPT, UPS, Toulouse, France}
\date{}
\maketitle
\begin{abstract}
We compute the leading-order inertial corrections to the instantaneous force acting on a rigid body moving with a time-dependent slip velocity in a linear flow field, assuming that the variation of the undisturbed flow at the body scale is much larger than the slip velocity between the body and the fluid. Motivated by applications to turbulent particle-laden flows, we seek a formulation allowing this force to be determined for an arbitrarily-shaped body moving in a general linear flow. 
 We express the equations governing the flow disturbance in a non-orthogonal coordinate system moving with the undisturbed flow and solve the problem using matched asymptotic expansions. The use of the co-moving coordinates enables the leading-order inertial corrections to the force to be obtained at any time in an arbitrary linear flow field. We then specialize this approach to compute the time-dependent force components for a sphere moving in three canonical flows: solid body rotation, planar elongation, and uniform shear. We discuss the behaviour and physical origin of the different force components in the short-time and quasi-steady limits. Last, we illustrate the influence of time-dependent and quasi-steady inertial effects by examining the sedimentation of prolate and oblate spheroids in a pure shear flow. 
\end{abstract}

\section{Introduction}
\label{intro}
\noindent The pioneering experiments by \cite{Segre1962a,Segre1962b} revealed that small neutrally-buoyant particles in a Poiseuille flow experience a lateral migration that concentrates them in an annulus with a well-defined radius. Almost simultaneously, \cite{Bretherton1962} proved that Stokes-type solutions for the disturbance induced by an axisymmetric body immersed in an arbitrary flow field cannot result in any lift force. These two discoveries, combined with the recent success of the matched asymptotic expansions technique (hereinafter abbreviated as MAE) in low-Reynolds-number flows \citep{Kaplun1957,Proudman57}, strongly stimulated the theoretical study of shear-induced inertial effects acting on rigid spherical or spheroidal bodies. In a seminal paper based on the use of an improved version of the MAE approach \citep{Childress64}, \cite{Saffman65} showed that a sphere translating along the streamlines of a simple shear flow with a non-zero slip velocity experiences a force that acts to move it perpendicular to the streamlines. His work, in which inertial effects due to the shear are assumed to dominate over those due to the slip between the undisturbed flow and the sphere, motivated a stream of theoretical and experimental studies during the next 50 years (see \cite{Stone00} for an overview covering the first 35 years). In particular, \cite{Harper68} worked out the case of a sphere translating in an arbitrary direction in a simple shear flow, determining inertial corrections to the drag in every direction as well as the components of the lift force. \cite{McLaughlin91} extended Saffman's approach to physical situations in which slip effects are of similar magnitude or even larger than those of the shear. His results revealed that inertial effects due to the slip velocity (i.e. Oseen-like effects) decrease the magnitude of the lift force compared to Saffman's prediction and may even change its sign when the shear becomes weak enough. \\
All aforementioned results only hold for solid spheres and quasi-steady conditions. They were generalized to spherical drops and bubbles by \cite{Legendre97} who showed in particular that the lift force on a spherical bubble with a vanishingly small viscosity is $(2/3)^2$ that on a solid sphere. Saffman's analysis was also extended to the case of a sinusoidally time-varying slip velocity by \cite{Asmolov99} who determined that the component of the lift force in phase with the slip decreases monotonically as the frequency of the excitation increases, whereas the out-of-phase component (which vanishes in the quasi-steady limit) goes through a maximum when this frequency is of the same order as the shear rate. An approximate transformation of these results to the time domain was achieved by \cite{Candelier07}, thus yielding the lift force for arbitrary time-dependent sphere translations along the streamlines of the shear flow. \vspace{2mm}\\
\noindent Besides the unidirectional shear flow, two other configurations involving bidirectional planar linear flows were also considered using the MAE approach, still under Saffman's assumptions. One of them corresponds to the case of a purely elongational flow. It was worked out by \cite{Drew78} who showed that a sphere translating in a direction that is not aligned with one of the principal axes of the strain experiences a  lift force, and hence tends to be deflected toward one of these axes. The second of these bidirectional configurations is the one where the sphere is immersed in a solid-body rotation flow, which actually involves two contrasting sub-cases. In the first of them, considered by \cite{Herron75}, the sphere is entrained by the flow (as in a centrifuge) and the slip velocity is most often in the radial direction, due to the density difference between the body and the fluid. Hence the `lift' force is circumferential and causes the sphere to lead the flow or to lag behind it, depending on the sign of the density difference. In contrast, in the second sub-case, the slip is oriented along the streamlines of the rotating flow and is such that the sphere appears fixed in the laboratory. This is the configuration obtained by releasing bubbles or rigid particles in a rotating cylinder with a horizontal axis \citep{VanNierop2007,Bluemink2010,Sauma2018}. This situation was considered by \cite{Gotoh90}. The connection between these two sub-cases, which yield different values for the inertial contributions to the hydrodynamic force, was clarified by \cite{Candelier08} who established the general form of this force encapsulating the two situations. More specifically, he showed that this force involves in general two distinct `history' terms accounting for the fact that in the first (resp. second) sub-case the sphere velocity is time-dependent when evaluated in the laboratory (resp. rotating) frame. \vspace{2mm}\\
In the above review, we did not discuss the influence of nearby walls on the lift force. Near-wall effects change the velocity profile of the undisturbed flow, induce slip between the body and the fluid even in the case of neutrally-buoyant particles, and modify the disturbance flow resulting from the presence of the latter. Obviously they are of primary importance to understand migration phenomena in pipe and channel flows, as well as in rotating containers. These problems involve regular or singular perturbations, depending on whether the wall-particle separation distance is smaller or larger than the distance at which inertial effects start to modify the structure of the disturbance. Several forms of Lorentz's reciprocal theorem and the MAE approach have been used to compute the lift force. We refer the reader to papers by \cite{Leal1980}, \cite{Hogg94} and \cite{Magnaudet2003} for reviews of the corresponding studies. \vspace{0mm}\\
 Coming back to unbounded linear flows, another series of studies employed the so-called `induced-force' method (hereinafter abbreviated as IF) as an alternative to the MAE approach, based on the formulation developed by \cite{Mazur74} to extend Fax\'en's formulae to a sphere undergoing an arbitrary  time-dependent motion in an inhomogeneous flow. In this method, an extra force is added to the Navier-Stokes equation to ensure that the slip velocity vanishes everywhere within the body, rendering the modified equation valid in the entire domain, both in the fluid and the body. This approach was first applied by \cite{Bedeaux87} to find the frequency-dependent inertial corrections to the force experienced by a sphere translating in a planar or an axisymmetric purely elongational flow. \cite{PerezMadrid90} then obtained the quasi-steady form of the friction tensor for the three canonical planar flow configurations discussed above. 
 While their result agreed with that of \cite{Herron75} in a solid-body rotation flow, the components of the resistance tensor obtained in the case of a pure shear flow differed from those determined by \cite{Harper68}. In particular the component corresponding to the Saffman's lift force was found to be approximately 2.3 times larger than predicted by the MAE approach \citep{Saffman1968, Harper68}. This issue was reconsidered by \cite{Miyazaki95a} who identified that a non-algebraic term was unduly neglected by \cite{Bedeaux87} and \cite{PerezMadrid90}, leading to erroneous results in the quasi-steady limit (except in the solid-body rotation case where this term does not contribute to the final result). Having dealt with this term through a transformation described later, \cite{Miyazaki95a}  could recover Saffman's prediction and conclude that the MAE and IF approaches yield identical results as expected. \cite{Miyazaki95b} employed the same technique to clarify the connection between the quasi-steady results established by \cite{Herron75} and \cite{Gotoh90} in a solid-body rotation flow.\vspace{2mm}\\
The above review indicates that the current knowledge regarding low-but-finite-Reynolds-number shear-induced lift forces acting on a sphere is quite satisfactory for both pure shear and solid-body rotation flows, for which the quasi-steady and frequency-dependent behaviours are now known. The situation is more uncertain as far as the elongational flow is concerned, since only the quasi-steady expression of the inertial corrections derived by \cite{Drew78} may potentially be correct in that case. Besides these three canonical configurations, no result has been established for more general linear flows resulting from an arbitrary combination of strain and rotation, neither in two nor in three dimensions. Once expressed in the proper eigen-basis, these flow fields depend on two and five independent parameters, respectively, in contrast with only one parameter (the shear, rotation, or strain rate) in the above three configurations. In such situations, inertial corrections cannot be obtained by linearly superposing expressions available for solid-body rotation and planar elongation, since the governing equation for the flow disturbance is nonlinear. This is obviously a major limitation on the route toward accurate predictions of particle motion in turbulent flows. Indeed, provided that the characteristic time and length scales over which the local structure of the carrying eddies varies are large enough compared with those involved in the disturbance, the background flow past the particle may be considered as linear and time-independent. This is why determining inertial corrections to the hydrodynamic force in this generic situation would represent a major step forward in the study of turbulent particle-laden flows.\vspace{2mm}\\
The chief technical difficulty in this class of problems is that, owing to the presence of a space-dependent term (the one that was overlooked by \cite{Bedeaux87} and \cite{PerezMadrid90}), the unsteady disturbance is governed by a set of coupled partial differential equations. This makes the solution particularly difficult to obtain. For the solid-body rotation, this difficulty is easily overcome by using a rotating reference frame, since the space-dependent term disappears in this frame \citep{Herron75,Miyazaki95b,Candelier08}. Based on this observation, it seems natural to seek a generic coordinate transformation that removes this term whatever the carrying flow. This is the backbone of the present work. More precisely, we express the unsteady disturbance problem in a system of moving non-orthogonal coordinates that follow the undisturbed flow. The disturbance is then determined by a set of ordinary differential equations in these co-moving coordinates, making the problem much easier to solve. Solving these equations and transforming back to the laboratory frame yields the desired inertial corrections irrespective of the nature of the linear carrying flow. Actually, this idea was already used by \cite{Miyazaki95a}, extending a technique developed by \cite{onuki1980} for a scalar field, but they employed it in a different way, namely by considering time-dependent wavenumbers in the Fourier transform of the disturbance equation.\vspace{2mm}\\
 \noindent A second major limitation of the results available in the aforementioned literature is that they were essentially derived for spherical bodies. However most rigid particles involved in flows of geophysical or engineering relevance are not spherical, which makes the determination of the lift force on non-spherical bodies moving in linear flows of particular relevance. The first step in that direction was taken by \cite{Harper68} who, by using the inertial corrections acting on a sphere translating in an arbitrary direction in a linear shear flow, found a way to evaluate these corrections for a spheroid with an arbitrary aspect ratio. However, a key assumption in their approach is that the disturbance is steady, whereas non-spherical bodies immersed in a shear flow rotate, thus inducing an unsteady disturbance. In particular, spheroidal bodies are known to tumble periodically in the zero-Reynolds-number limit \citep{Jeffery22}. Because of this periodic angular motion, a relevant calculation of the lift force acting on spheroidal particles must take unsteady effects into account. \\
\noindent The leading-order inertial corrections to the loads experienced by a rigid body depend linearly on the force and torque acting on it in the zero-Reynolds-number limit, and this force and torque depend linearly on the relative velocity and rotation rate between the body and fluid through resistance tensors entirely determined by the body geometry \citep{Brenner63,Brenner64a,Brenner64b, Happel83}. How does the shape of the body enter the MAE and IF approaches? In the former, inertial corrections are entirely governed by the uniform component of the residual disturbance  in the `outer' domain \citep{Kaplun1957,Proudman57}. Moreover, as first recognized by \cite{Childress64}, the presence of the body enters only through a point-source term in the `outer' problem. Hence, as far as the resistance tensors are known in the zero-Reynolds-number limit, it is straightforward to extend results derived for spherical bodies to arbitrarily-shaped bodies. In contrast, it is not obvious to infer how the IF method can be generalized to such bodies, and to the best of our knowledge this has not been achieved yet. Indeed this method requires the slip between the body and fluid to vanish everywhere within the inner domain bounded by the actual body surface. Hence resistance tensors are not involved explicitly, making the prescription of the `induced force' non trivial as soon as a non-spherical shape is considered. This is why the MAE technique appears to be much more suitable for bodies of general shape.\vspace{2mm}\\
\noindent The primary goal of our work is to determine, using the MAE technique, how the instantaneous force and torque on an arbitrarily-shaped body are affected by small inertia effects in a general quasi-steady linear flow. The present paper describes our approach in its full generality, but the applications presented hereinafter only concern the usual three canonical planar flows, and in addition the sedimentation of spheroidal particles in a shear flow. In that respect, the present contribution only represents a first step toward the `Holy Grail' of the prediction of finite inertial effects acting on particles immersed in a turbulent flow.

\noindent The paper is organized as follows. Section \ref{formul} outlines the formulation of the problem, stating in particular the asymptotic conditions under which the solution is sought, establishing the corresponding disturbance flow problem, and deriving the form of unsteady force and torque corrections in terms of the solution of this problem.
Section \ref{sec:coords} describes how to solve the disturbance problem by introducing a moving non-orthogonal coordinate system. This reduces the initial problem to a set of ordinary differential equations in Fourier space, and provides the general structure of the corresponding solution, first in Fourier space, then in the physical space in the form of a tensorial convolution kernel. The technical steps leading to the solution, which involve in particular the use of Magnus expansions, are detailed in appendix \ref{AppendixA}. In section \ref{bidir}, we compute explicitly the kernel in the solid-body rotation and planar elongational flows, and examine its various components at both short and long time after the flow has started impulsively. Section \ref{sec:kernel} considers specifically the case of the one-directional linear shear flow. After computing the kernel, we discuss the physical mechanisms that govern the evolution of its various components. In section \ref{sec:sedimentation}, the results obtained in the previous sections are finally applied to the sedimentation of an arbitrary-shaped body immersed in a linear flow. After general results are established, the focus is put on the sedimentation of spheroids in a uniform shear; light is shed on the role of the time-dependent inertial corrections by comparing the evolution of the slip velocity of the spheroid predicted under various approximations. Section \ref{sec:conc} summarizes the main findings of this investigation and draws some perspectives for future work.

\section{Formulation of the problem}
\label{formul}
\subsection{Statement of the general problem}
\label{statpb}
\label{sec:problem}
\begin{figure} 
\mbox{}\\
\begin{center}
\begin{overpic}[width=0.5\textwidth]{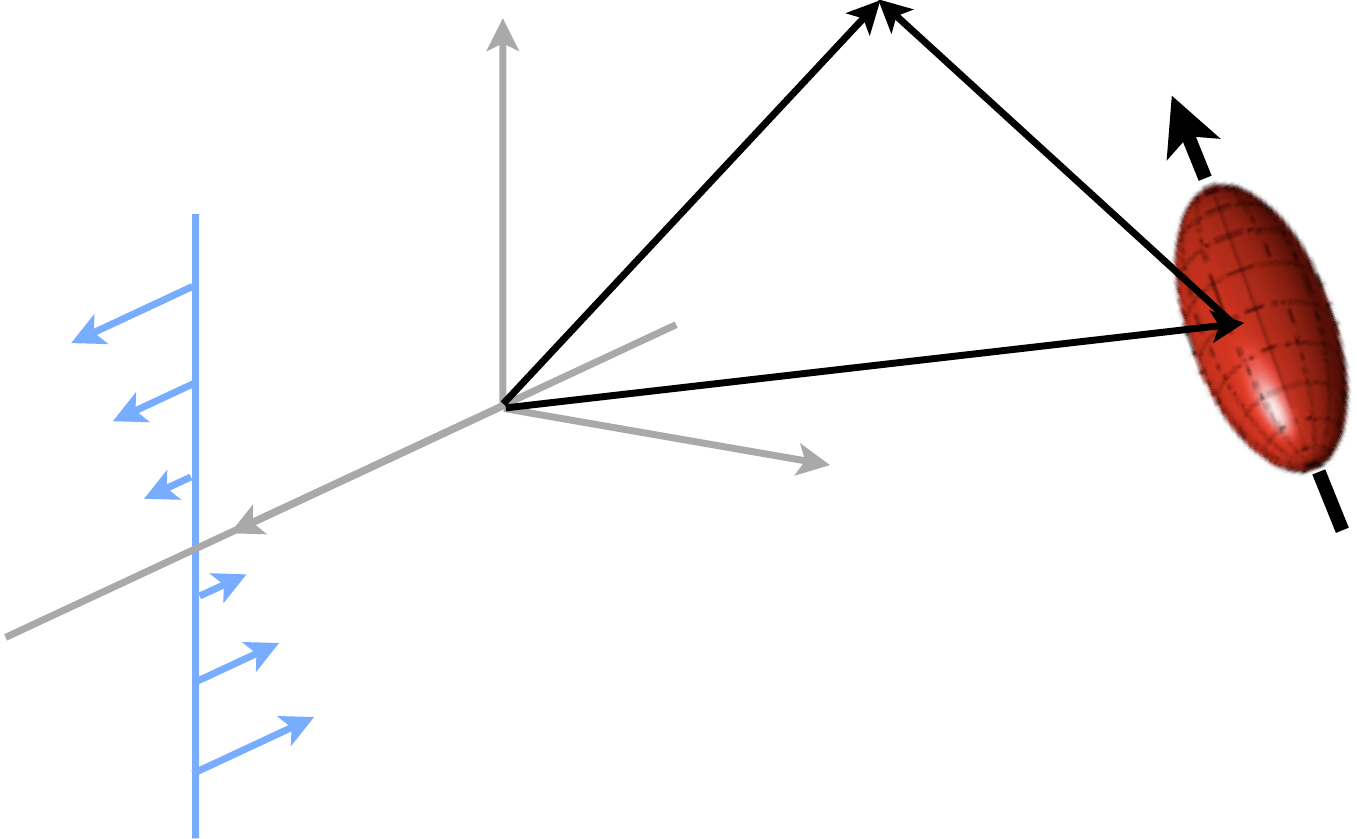}
\put(31,60){$\vec{e}_3$}
\put(32,33){$\vec{o}$}
\put(20,20){$\vec{e}_1$}
\put(61,24){$\vec{e}_2$}
\put(8,49){$\vec{U}^\infty(\vec{x})$}
\put(70,31){$\vec{x}_p$}
\put(54,45){$\vec{x}$}
\put(69,49){$\vec{r}$}
\put(81,55){$\vec{n}$}
\end{overpic}
\end{center}
\caption{A particle is moving in a steady linear flow. The body shown in this sketch is a spheroid (with symmetry vector $\vec{n}$), but the method applies to arbitrary body shapes.}
\label{Diph_fig1}
\end{figure}
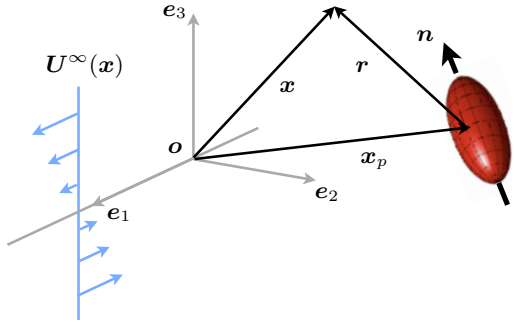
Figure \ref{Diph_fig1} illustrates the generic problem considered here, where we wish to determine the force  $\vec{f}$  on a small rigid body of arbitrary shape moving in a  linear flow.  The undisturbed fluid velocity
has the general form
\begin{equation}
\vec{U}^\infty(\vec{x}) =  \ma{A} \cdot \vec{x}\:,
\label{eq:v_infty}
\end{equation}
where $\ma{A}$ is a tensor with time- and space-independent components. 
\\
To describe the dynamics of the body, the force and the torque acting on it must be determined. The position of the body centre in the laboratory frame is defined by the vector $\vec{x}_p(t)$, while its translational velocity and angular velocity are denoted by $\dot{\vec{x}}_p(t)$ and $\boldsymbol{\omega}_p(t)$, respectively.  The body density is assumed to be uniform so that its centre of mass coincides with its geometrical centre. In the presence of gravity, the force $\vec{f}$ on a non-neutrally buoyant body immersed in an arbitrary flow is 
\begin{equation}
\vec{f}  =   m_p \:\vec{g}+ \int_{\mathcal{S}_p} \ma{\Sigma}^\infty \cdot \mbox{d}\vec{s}  +  
\underbrace{\int_{\mathcal{S}_p} \ma{\Sigma}' \cdot \mbox{d}\vec{s}}_{=\vec f'} \:,
\label{eq_force}
\end{equation}
where $m_p$ is the mass of the body and $\vec g$ denotes gravity.
Similarly, the torque $\vec{\tau}$ with respect to the body centroid is 
\begin{equation}
\vec{\tau}  =    \int_{\mathcal{S}_p} \vec{r} \times \ma{\Sigma}^\infty \cdot \mbox{d}\vec{s}  +  
\underbrace{\int_{\mathcal{S}_p} \vec{r} \times \ma{\Sigma}' \cdot \mbox{d}\vec{s}}_{=\vec{\tau}'} \:, \quad \mbox{where} \quad \vec{r} = \vec{x} - \vec{x}_p. 
\label{eq_torque}
\end{equation}
The integrals are over the body surface $\mathcal{S}_p$, $\mbox{d}\vec{s} $ is the outward normal surface element, $\ma{\Sigma}^\infty$ and $\ma{\Sigma}'$ are the stress tensors associated with the undisturbed flow and the disturbance, respectively. The last terms in the right-hand side of (\ref{eq_force}) and (\ref{eq_torque}) define the disturbance force, $\vec f'$, and the disturbance torque, $\vec{\tau}'$, respectively. 

Since the undisturbed flow is known, the integrals involving $\ma{\Sigma}^\infty$ in (\ref{eq_force}) and in (\ref{eq_torque}) are readily evaluated by using  Stokes' theorem together with the fact that the undisturbed flow is a solution of the Navier-Stokes equation \citep[e.g.][]{Gatignol83}. To  compute the disturbance force and torque, the stress tensor of the disturbance flow must be determined, requiring
in principle to solve the Navier-Stokes equations for the disturbance velocity $\vec w(\vec r, t)=\vec U(\vec r+\vec x_p,t)-\vec U^\infty(\vec r+\vec x_p,t)$.
In the following, we assume that this solution is known in the quasi-steady creeping-flow limit where no inertia effects are considered. In this limit, the disturbance force and torque may be expressed in the form \citep{kim1991}
\begin{equation}
\left[
\begin{array}{c}
\vec{f}'^{(0)}(t)\\
\vec{\tau}'^{(0)}(t)\\
\end{array}
\right]=
- \mu \left[
\begin{array}{cc}
\ma{M}_{1}(t) & \ma{M}_{2}(t) \\
\ma{M}_{2}^{\sf T}(t) & \ma{M}_{3}(t) \\
\end{array}
\right] \cdot \left[
\begin{array}{c}
\dot{\vec{x}}_p - \vec{U}^{\infty}\\
\vec{\omega}_p - \vec{\Omega}^{\infty}\\
\end{array}
\right] -\mu  \left[
\begin{array}{c}
\ma{N}_1(t) : \ma{S}^\infty\\
\ma{N}_2(t) : \ma{S}^\infty\\
\end{array}
\right]\,.
\label{eq_wrench}
\end{equation}
Here $\mu$ is the dynamic viscosity of the fluid, $\ma{S}^\infty \equiv \frac{1}{2}\ \left( \ma{A} + \ma{A}^{\sf T}\right)$ is the symmetric part of the strain-rate tensor $\ma{A}$ (throughout the paper, $^{\sf T}$ denotes the transpose), and $\vec{\Omega}^{\infty} \equiv \frac{1}{2}\boldsymbol{\nabla} \times \vec{U}^\infty$ is half the vorticity of the undisturbed flow.  The $\ma{M}_i$ are the second-order resistance tensors, while the $\ma{N}_i$  are third-order tensors. The components of $\ma{M}_i$ and of $\ma{N}_i$ in the principal basis of the body are assumed to be known. In the laboratory frame, these components depend on time, since they depend on the instantaneous orientation of the body. 

\subsection{Dimensionless parameters and variables}
In what follows, distances are normalized by a characteristic body length, $a$, so that  
$ \vec r = a \vec r^\prime$, and translational velocities are normalized by the typical order of magnitude of the body's slip velocity, $u_c$, so that  $\vec U =  u_c\vec U^\prime$. 
Consequently pressures are normalized in the form $p = (\mu u_c/a)p'$. Components of $\ma{A}$ are normalized by a characteristic velocity gradient, $s$, defined as $s= \sqrt{(\ma{A} : \ma{A})/2}$ in the bidirectional flows considered in section \ref{bidir}, and as $s= \sqrt{\ma{A} : \ma{A}}$ in the one-directional shear flow on which section \ref{sec:kernel} focuses. The two different normalizations ensure that $s$ is the magnitude of the velocity gradients in both cases. In the absence of a condition on the overall torque, the body rotation rate may be prescribed arbitrarily, so that it provides an independent time scale, ${\omega}_p^{-1}$. Last, it is necessary to introduce the characteristic time $\tau_c$ over which the relative translational and rotational velocities may vary at the body surface, so that the dimensionless time $t'$ is  $ t^\prime =  t/\tau_c$. \\
With these definitions, the problem is governed by four dimensionless numbers, namely 
\begin{equation}
\mbox{Re}_s = \frac{a^2 s}{\nu}\,,\quad \mbox{Re}_p = \frac{a u_s}{\nu}\,,\quad \mbox{Re}_\omega = \frac{a^2\omega_p}{\nu}\ \mbox{and}\quad  \mbox{Sl} = \frac{1}{s {\tau_c}} \:,
\end{equation}
where $\mbox{Re}_s$, $\mbox{Re}_p$ and $\mbox{Re}_\omega$ are the shear, slip and rotation Reynolds numbers, respectively, $\mbox{Sl}$ is the Strouhal number characterizing the unsteadiness of the problem and $\nu=\mu/\rho_f$ is the kinematic viscosity. 
In the remainder of the paper, the primes are dropped but it must be understood that all equations and quantities are dimensionless.
In these variables, the equations governing the disturbance flow, expressed in a frame translating with the body, read
\begin{equation}
\mbox{Re}_s\mbox{Sl} \;\partial _t\vec{w}\big|_{\vec{r}}  + \mbox{Re}_s \big[\ma{A} \cdot \vec{w} + 
(\ma{A} \cdot \vec{r})\cdot \tens{\nabla}  \vec{w} \big]  +\mbox{Re}_p \big[-  \vec{u}_s \cdot \tens{\nabla} \vec{w}  +  \vec{w} \cdot \tens{\nabla} \vec{w} \big] = -  \boldsymbol{\nabla} p +\boldsymbol{\nabla}^2 \vec{w}\:,
\label{eq_w_1_2}
\end{equation}
subject to the incompressibility condition
\begin{equation}
\boldsymbol{\nabla} \cdot \vec{w}= 0\:,
\label{continuity_eqw1_2}
\end{equation}
and to the boundary conditions
\begin{equation} 
 \vec{w} \rightarrow 
\vec{0}\quad \mbox{for}\:\: |\vec{r}|\to \infty
\quad \mbox{and} \quad 
\vec{w} = \vec{u}_s + \frac{\mbox{Re}_\omega}{\mbox{Re}_p}\vec{\omega}_p \times \vec{r} - \frac{\mbox{Re}_s}{\mbox{Re}_p}(\vec{\Omega}^\infty\times \vec{r}+\ma{S}^\infty \cdot \vec{r}) \: \:\:\mbox{for}\:
\quad \vec{r} \in \mathcal{S}_p  \:.
\label{cl1_2}
\end{equation}
In (\ref{eq_w_1_2}) and (\ref{cl1_2}), the gradients are spatial derivatives with respect to $ \vec{r} = \vec{x} - \vec{x}_p$, the time derivative is evaluated at fixed $\vec{r}$, and we have introduced
the slip velocity $\vec{u}_s=\dot{\vec{x}}_p - \vec{U}^\infty(\vec{x}_p) $. 
We consider the problem in the Saffman limit where all three Reynolds numbers are small and satisfy the ordering conditions (the origin of which is discussed later in this paragraph)
\begin{equation}
\mbox{Sup}({\mbox{Re}_\omega,\,\mbox{Re}_s})\sqrt{\mbox{Re}_s}\ll \mbox{Re}_p \ll \sqrt{\mbox{Re}_s}  \ll 1 \:.
\label{Saffman_condition}
\end{equation}
We further assume that 
\begin{equation}
\mbox{Sl} \ll \frac{1}{\mbox{Re}_s}\:,
\label{eq_Strouhal}
\end{equation}
which allows unsteadiness to be large, but not `too' large. In particular, $\tau_c$ must in principle be much larger than the viscous time $a^2/\nu$.
Conditions (\ref{Saffman_condition}) and (\ref{eq_Strouhal}) significantly simplify the problem which can then be treated $via$ a perturbation approach. Close to the body, the disturbance velocity is then well approximated by the  quasi-steady Stokes solution $\vec{w} \sim 1/r$. Thanks to this scaling form, it is readily found that beyond the so-called Saffman length defined as 
\begin{equation}
|{\bf{r}}| \sim \frac{1}{\sqrt{\mbox{Re}_s}}\equiv\ell_s\:,
\end{equation}
 terms proportional to ${\mbox{Re}_s}$ in (\ref{eq_w_1_2}) become negligibly small compared to the other terms in the disturbance equation. The part of the solution that stems from the term $(\vec{\omega}_p - \vec{\Omega}^\infty) \times \vec{r} - \ma{S}^\infty \cdot \vec{r}$ in (\ref{cl1_2}) consists of a combination of rotlets and stresslets  which decay as $1/r^2$ close to the body \citep{Lamb45}. Hence, at distances of the order of $\ell_s$, the second and third terms on the right-hand side of (\ref{cl1_2}) induce disturbances that are respectively of ${\mathcal{O}}({\mbox{Re}}_\omega{\mbox{Re}}_s^{1/2}{\mbox{Re}}_p^{-1})$  and ${\mathcal{O}}({\mbox{Re}}_s^{3/2}{\mbox{Re}}_p^{-1})$ compared to that of the first term. Hence, the first inequality in (\ref{Saffman_condition}) guarantees that, beyond the Saffman length, these contributions are negligibly small. With this set of assumptions, the disturbance problem in the far field thus reduces to
 \begin{equation}
 \mbox{Re}_s \mbox{Sl}\,  \displaystyle{\partial_t \vec{w}}  \big|_{{\vec{r}}} +  \mbox{Re}_s  \Big( \ma{A} \cdot \vec{w} + 
(\ma{A} \cdot \vec{r})\cdot \tens{\nabla}  \vec{w} \Big)  = -  \boldsymbol{\nabla} p +\boldsymbol{\nabla}^2 \vec{w}\:,
\label{eq_w_1_3}
\end{equation}
\begin{equation}
\boldsymbol{\nabla} \cdot \vec{w}= 0\:,
\label{continuity_3}
\end{equation}
\begin{equation} 
 \vec{w} \rightarrow 
\vec{0}\: \: \mbox{for}\quad  |\vec{r}|\to \infty
\quad \mbox{and} \quad 
\vec{w} = \vec{u}_s \: \:\:\mbox{for} 
\quad \vec{r} \in \mathcal{S}_p \:.
\label{cl1_3}
\end{equation}

\subsection{Asymptotic solution}
\label{asympt}
As discussed in the introduction, the MAE approach has been extensively employed to determine how small inertia effects alter the force acting on a small rigid or deformable body since the pioneering studies of \cite{Kaplun1957} and \cite{Proudman57}. The standard method used in the presence of a non-uniform background flow was devised by \cite{Childress64} and \cite{Saffman65}. Specificities of this method are that the body is replaced by a point-force (through a Dirac-delta function with appropriate strength and direction) in the far-field equation of the disturbance, and that the matching is performed in Fourier space. In all studies reviewed in \S\ref{intro}, with the 
 exception of the work by \cite{McLaughlin91}, expansions were performed with respect to the small parameter 
\begin{equation}
 \epsilon =\ell_s^{-1}= \mbox{Re}_s^{1/2}\:.
\end{equation} 
This is also the key small parameter here. However, most studies to date based on the MAE approach considered either the quasi-steady approximation $\mbox{Sl}\rightarrow0$ or harmonic variations of the slip velocity \citep{Asmolov99}. Here in contrast, unsteady effects corresponding to arbitrary evolutions are considered up to ${\mathcal{O}}(\epsilon^{2-n})$ with $0\le n<2$, together with ${\mathcal{O}}(\epsilon^2)$ shear-induced inertial effects. As will become apparent soon, the additional difficulty resulting from such unsteady effects is that the governing equation for the velocity disturbance becomes a partial differential equation with respect to time and space, which greatly complicates the determination of its solution. This is why, to the best of our knowledge, only \cite{Candelier08} considered such arbitrary evolutions in the specific case of a solid-body rotation flow, where this additional difficulty is overcome by a simple change of reference frame.\\

As discussed above, under conditions (\ref{Saffman_condition}), the overlap between the inner and outer regions of the perturbation problem arises at distances from the body of the order of the Saffman length,
$\ell_s=\epsilon^{-1}$.\vspace{2mm} \\
In the inner region, $|\vec{r} | \ll 1/\epsilon$, the disturbance is sought  in the form of a regular expansion in powers of $\epsilon$, namely
\begin{equation}
\vec{w}_{\mbox{\scriptsize in}} = \vec{w}_{\mbox{\scriptsize in}} ^{(0)} + \epsilon \: \vec{w}_{\mbox{\scriptsize in}} ^{(1)} + \ldots \quad \mbox{and} \quad  p_{\mbox{\scriptsize in}}  = {p}_{\mbox{\scriptsize in}} ^{(0)} + \epsilon \: {p}_{\mbox{\scriptsize in}} ^{(1)} + \ldots \:
\end{equation}
The inner solution, $\vec w_{\rm in}$, satisfies the required boundary condition at the body surface but fails in the outer region $|\vec{r}| \gg 1/\epsilon$ since it tends to produce inertial contributions of larger magnitude than the viscous term \citep{Oseen1910,VanDyke1978}.
Therefore, inertial and viscous terms must both be considered in the outer region. Following \cite{Childress64}, the effect of the inner boundary condition is replaced by a point force, $\vec{f}^{(0)}$, weighted by the three-dimensional delta function, $\delta(\vec{r})$ . Hence, assuming provisionally $\mbox{Sl}={\mathcal{O}}(1)$, the disturbance flow is governed by
\begin{equation}
\epsilon^2 \big[\partial _t\vec{w} \big|_{\vec{r}}  +  \ma{A} \cdot \vec{w} + 
(\ma{A} \cdot \vec{r})\cdot \tens{\nabla}  \vec{w} \big]    = - \boldsymbol{\nabla} p +
 \boldsymbol{\nabla}^2 \vec{w} +\vec{f}^{(0)} \delta(\vec{r})\:,
\label{eq_w_adime}
\vspace{3mm}
\end{equation}
subject to the divergence-free condition (\ref{continuity_3}), and to the vanishing condition $\vec{w} \rightarrow\vec{0}$ for $|\vec{r} |\rightarrow\infty$.
To leading order, $\vec{f}^{(0)} $ is just the leading-order force exerted by the body on the fluid, i.e. the opposite of the Stokes force acting on the body in the linear flow as given by (\ref{eq_wrench}). 
Thus we write
\begin{equation}
\vec{f}^{(0)}(t) \equiv - \vec{f}'^{(0)}(t) = \ma{M}_1(t) \cdot \vec{u}_s(t) + \ma{M}_2(t) \cdot \vec{\omega}_s(t) + \ma{N}_1(t): \ma{S}^\infty\:.
\label{eq_f_0}
\end{equation}

\noindent As mentioned in \S \ref{statpb}, the components of  $\ma{M}_1$, $\ma{M}_2$ and $\ma{N}_1$ generally depend upon time.
The situation greatly simplifies for a sphere, for which $\ma{M}_1 =  6 \pi \ma{1}$ and  $\ma{M}_2=\ma{N}_1=\ma{0}$, where $\ma{1}$ is the identity (Kronecker) tensor.\\
Equation (\ref{eq_w_adime}) was written under the assumption $\mbox{Sl}={\mathcal{O}}(1)$. However, following (\ref{eq_Strouhal}), we stress again that the solutions derived throughout this paper are valid even for larger Strouhal numbers, $\mbox{Sl}={\mathcal{O}}(\epsilon^{-n})$ with $n\ge0$, provided that $n<2$.  \vspace{2mm}\\
Fourier transforming (\ref{eq_w_adime}) yields
\begin{equation}
\epsilon^2\left[ \displaystyle{\partial_t \hat{\vec{w}}} \big|_{\vec{k}}  +  \ma{A} \cdot \hat{\vec{w}} -
 \vec{k} \cdot \ma{A} \cdot  \hat{\boldsymbol{\nabla}} 
\hat{\vec{w}}  \right]
= - \mathtt{i} \vec{k} \:
 \hat{p} - k^2 \:\hat{\vec{w}} +\vec{f}^{(0)}\:,
\label{eq_w_hat}
\end{equation} 
where $\mathtt{i}^2=-1$, $k^2 = \vec{k}\cdot \vec{k}$, $\hat{\boldsymbol{\nabla}}$ denotes the gradient with respect to $\vec{k}$, and the direct and inverse Fourier transforms are respectively defined as
\begin{align}
\vec{\hat{w}}  = \int%%_{\mathbb{R}^3} 
\vec{w}(\vec{r},\:t) \:\exp(- \mathtt{i} \:\vec{k}\cdot \vec{r} ) \:\mbox{d}^3 \vec{r}
%% \quad \mbox{where} \quad \mathtt{i}^2=-1
\quad\mbox{and}\quad
\vec{w}  = \frac{1}{8\pi^3}\int%%_{\mathbb{R}^3} 
\vec{\hat{w}}(\vec{k},\:t) \exp(\mathtt{i} \:\vec{k}\cdot \vec{r} )\, \mbox{d}^3 \vec{k}\:.
\end{align} 

\noindent The solution of the outer problem (\ref{eq_w_hat}) may be expanded in terms of generalized functions of $\vec{k}$ \citep{Meibohm2016} with respect to $\epsilon$, i.e.
\begin{equation}
\hat{\vec{w}}(\vec{k},\,t) = \hat{\vec{\mathcal{T}}}^{(0)}(\vec{k},\,t) + \epsilon  \hat{\vec{\mathcal{T}}}^{(1)}(\vec{k},\,t) + O(\epsilon^2)\:.
\end{equation}
When transformed back to the physical domain, ${\mathcal{T}}^{(0)}$ and ${\mathcal{T}}^{(1)}$ read 
\begin{equation}
\vec{{\mathcal{T}}}^{(0)}(\vec{r},\,t) = \frac{1}{8 \pi} \left( \frac{\ma{1}}{|\vec{r}|}+\frac{\vec{r} \vec{r}}{|\vec{r}|^3} \right) \cdot \vec{f}^{(0)}(t)
\label{eq_T0}
\end{equation}
and 
\begin{equation}
\vec{{\mathcal{T}}}^{(1)}(t) =\frac{1}{8\pi^3} \int \big[\hat{\vec{w}}(\vec{k},\,t)|_{\epsilon=1}  -  \hat{\vec{\mathcal{T}}}^{(0)}(\vec{k},\,t)  \big]  \mbox{d}^3 \vec{k} \:.
\label{eq_T1}
\end{equation}
The leading-order term, ${\vec{\mathcal{T}}}^{(0)}(\vec{r},\,t)$, is the solution of the outer problem (\ref{eq_w_hat})  for $\epsilon=0 $, that is the Stokeslet solution. The contribution $\hat{\vec{w}}(\vec{k},\:t)|_{\epsilon=1}$ in (\ref{eq_T1}) is the solution of the outer problem (\ref{eq_w_hat}) for $\epsilon=1$. Remarkably, $\vec{\mathcal{T}}^{(1)}(t)$ turns out to be $\vec{r}$-independent; it therefore defines a uniform velocity in the far field.   
Equations (\ref{eq_T0}) and (\ref{eq_T1}) provide the boundary conditions to be satisfied by the inner solution for $|{\bf{r}}| \sim\epsilon^{-1}$. \vspace{2mm}\\
The inner and outer solutions must match at  $|{\bf{r}}| \sim \epsilon^{-1}$. Hence the lowest-order term
in the inner expansion, $\vec{w}_{\mbox{\scriptsize in}} ^{(0)}$, corresponds to the solution of Stokes equation satisfying
\begin{equation}
\vec{w}_{\mbox{\scriptsize in}} ^{(0)}  = \vec{u}_s \quad \mbox{for} \quad \vec{r} \in \mathcal{S}_p\: \quad \mbox{and} \quad \lim_{|\vec{r}| \to \infty} \vec{w}_{\mbox{\scriptsize in}} ^{(0)} \sim {\vec{\mathcal{T}}}^{(0)}(\vec{r},\,t)\:.
\end{equation}
The second term in the inner expansion, $\vec{w}_{\mbox{\scriptsize in}} ^{(1)}$, is also a solution of the Stokes equation, but now with boundary conditions
\begin{equation}
\vec{w}_{\mbox{\scriptsize in}} ^{(1)}  = \vec{0} \quad \mbox{for} \quad \vec{r} \in \mathcal{S}_p\: \quad \mbox{and} \quad \lim_{|\vec{r}| \to \infty} \vec{w}_{\mbox{\scriptsize in}} ^{(1)} \sim {\vec{\mathcal{T}}}^{(1)}(t)\:.
\end{equation}
Determining  $\vec{w}_{\mbox{\scriptsize in}}^{(1)}$ is thus equivalent to solving the Stokes flow problem about the body kept fixed in a uniform stream. Hence, considering inertia effects is equivalent to considering an additional uniform flow at infinity, so that one readily concludes that 
the (dimensionless) disturbance force in (\ref{eq_force}) 
reads to order $\epsilon$
\begin{equation}
\vec{f}' =   -  \ma{M}_1(t) \cdot\vec{u}_s  -  \ma{M}_2(t) \cdot\vec{\omega}_s - \ma{N}_1(t): \ma{S}^\infty+ \epsilon \ma{M}_1(t) \cdot    \Big\{ \frac{1}{8\pi^3} \int_{\mathbb{R}^3} \big[\hat{\vec{w}}(\vec{k},\:t)|_{\epsilon=1}  -  \hat{\vec{\mathcal{T}}}^{(0)}(\vec{k},t)  \big] \mbox{d}^3 \vec{k}\Big\}\:,
\label{eq:force}
\end{equation} 
while the disturbance torque in (\ref{eq_torque}) reads 
\begin{equation}
\vec{\tau}' =   -  \ma{M}_2^{\sf T}(t) \cdot\vec{u}_s  -  \ma{M}_3(t) \cdot\vec{\omega}_s - \ma{N}_2(t): \ma{S}^\infty + \epsilon \ma{M}_2^{\sf T}(t) \cdot    \Big\{ \frac{1}{8\pi^3} \int_{\mathbb{R}^3} \big[\hat{\vec{w}}(\vec{k},\:t)|_{\epsilon=1}  -  \hat{\vec{\mathcal{T}}}^{(0)}(\vec{k},t)  \big] \mbox{d}^3 \vec{k}\Big\}\:,
\label{eq:torque}
\end{equation} 
These are the desired expressions for the unsteady force and torque acting upon an arbitrarily-shaped rigid body moving in a general linear flow.\vspace{2mm}\\
\noindent The chief difficulty in finding $\hat{\vec w}$  results from the advective term  
\begin{equation}
\label{inhom}
(\ma{A}\cdot \vec{r}) \cdot \tens{\nabla} \vec{w}
\end{equation}
on the left-hand side of (\ref{eq_w_adime}). Although this term is linear, it is inhomogeneous in the sense that it explicitly depends on the $\vec{r}$ vector, yielding partial derivatives with respect to $\vec{k}$ in (\ref{eq_w_hat}). This is what renders the determination of the solution technically difficult, even in the quasi-steady approximation, except under particular circumstances. One of these is the case where all spatial derivatives in (\ref{eq_w_hat}) are with respect to the same component of $\vec{k}$, as in Saffman's original problem. As explained in the introduction, the key idea here is to obtain a general solution to the disturbance problem by removing the  inhomogeneous advective term (\ref{inhom}) with the aid of a non-orthogonal coordinate system that moves and deforms with the undisturbed flow, so as to reduce this problem to a set of ordinary differential equations with respect to time that are much more easily solved.

\section{Solution strategy}
\label{sec:coords}
\subsection{Time-dependent non-orthogonal coordinates}
Although (\ref{eq_w_adime}) is written in a reference frame translating with the position of the body centroid, $\vec x_p(t)$, the unit vectors are those of the laboratory frame. Components $r^i$ of the vector $\vec r$
in the corresponding basis with unit vectors $\vec{e}_i$ define a rectilinear orthogonal coordinate system. 
Alternatively, one may introduce a new system with coordinates  $R^i$ such that 
\begin{equation}
r^i = {F^i}_j R^j \:.
\label{eq:transf}
\end{equation}
Here the ${F^i}_j(t)$ are the time-dependent components of a  transformation matrix, $\ma{F}(t)$, and summation is implied on repeated indices. We assume $\mbox{det}(\ma{F})\neq 0$, so that the relation (\ref{eq:transf}) between $r^i$ and $R^i$ can be inverted. 
Since (\ref{eq:transf}) is linear and the ${F^i}_j$ do not depend on the variable $\vec r$,  the $R^i$ coordinate system remains rectilinear. In contrast, these coordinates are generally non orthogonal. We introduce the time-dependent basis, $\vec{E}_i$, associated with these coordinates as
\begin{equation}
\vec{E}_i(t) = \ma{F} (t)\cdot \vec{e}_i\,.
\label{eq:g_i}
\end{equation}
In what follows we adopt the convention that components of vectors and  tensors expressed in the $\vec{E}_i$ and $\vec{e}_i$ bases are denoted with uppercase and lowercase letters, respectively. For instance, we  write 
$\vec{w} = w^i(r^j ,\: t) \vec{e}_i  = W^i(R^j,\:t) \vec{E}_i$. All vectors are expressed in contravariant form, and their components are thus denoted with upper indices. Because they contract with such vectors, several second-order tensors need to be expressed in covariant or mixed form, thus involving lower indices; e.g. ${F^i}_j$ in (\ref{eq:transf}). In appendix \ref{AppendixA} we show that the partial time derivatives of any vector at fixed $r^i$ and fixed $R^j$ components are related through
\begin{equation}
\frac{\partial \vec{w}}{\partial t} \Big|_{r^i} = \frac{\partial (W^i(R^j,\:t) \vec{E}_i )}{\partial t} \Big|_{R^j} 
- \vec{v} \cdot \boldsymbol{\nabla}  \vec{w} \quad\mbox{with}\quad
\vec{v} \equiv \frac{\partial  \vec{r}}{\partial t} \Big|_{R^j} \:.
\label{eq:derivative_transform}
\end{equation}
The goal is now to express (\ref{eq_w_adime}) with respect to coordinates $R^j$ in such a way that the $\vec v$-term cancels the inhomogeneous advective
contribution $(\ma{A}\cdot \vec{r}) \cdot \boldsymbol{\nabla} \vec{w}$. This is achieved by setting
\begin{equation}
\vec{v} =  \ma{A}\cdot \vec{r} \:.
\label{eq:V_c2}
 \end{equation}
It then follows from (\ref{eq:transf}) that  
  $\ma{F}(t) $ must satisfy
\begin{equation}
 \frac{\mbox{d}\ma F}{\mbox{d}t} = \ma{A} \cdot \ma{F} \quad\mbox{with}\quad \ma{F}(0) = \ma{1}\,.
 \label{eq:dot_F}
 \end{equation}
 The initial condition ensures that $r^i = R^i$ at $t=0$. In continuum mechanics, $\ma F$ is the deformation gradient tensor mapping an infinitesimal vector $\mbox{d} \vec{x}$ corresponding to the initial configuration onto another infinitesimal vector $\mbox{d}\vec{X}$ corresponding to the deformed configuration; $\ma F^{\sf T}\cdot \ma F$ is referred to as the Cauchy-Green tensor \citep{Truesdell1965,Eringen1967}. It provides the square of the local change in distances due to deformation, since $\mbox{d}\vec{X}\cdot\mbox{d}\vec{X}=d\mathbf {x} \cdot(\ma F^{\sf T}\cdot \ma F) \cdot d\mathbf {x}$. In fluid mechanics, $\ma F$  arises in characterizing the time-dependent orientations of rod-like particles advected in turbulence \citep{Wil09,Voth16}; in this case, $\ma{F}$ maps the initial rod orientation onto the final one.\\ 
As  $\ma{A}$ is time-independent, the solution to (\ref{eq:dot_F}) is merely
\begin{equation}
\ma{F} = \ma{E}\mbox{xp}(\ma{A}\:t)\:.
\label{eq:Exp_A}
\end{equation}
Now it remains to determine how the derivatives in  (\ref{eq_w_adime}) transform. Noting that the transformation (\ref{eq:transf}) depends linearly on the $R^i$, and using the fact that
$\frac{\mbox{d}\vec{E}_i}{\mbox{d} t } =  \ma{A} \cdot \vec{E}_i$, it is readily found that the $i$th component of (\ref{eq_w_adime}) transforms into
\begin{equation}
\frac{\partial W^i}{\partial t}\Big|_{R^i} +   \: 2 \: \ma{A}^i_{\:j} W^j   =  - \ma{R}^{ij}(t) \frac{\partial P}{\partial R^j} +  \ma{R}^{jk} (t) \frac{\partial^2 W^i}{\partial R^j \partial R^k} + {F^{(0) i}}(t)\delta(R^i\vec{E}_i)\:,
\label{eq_NS4} 
\end{equation}
where $\ma{R}^{ij}(t) = \delta^{\ell k} {{(\ma{F}^{-1})}^i}_ \ell {({\ma{F}^{-1})}^j}_k$, $\delta^{\ell k}$ denoting the Kronecker symbol. The $\ma{R}^{ij}$ are the components of the inverse of the metric tensor with components $\ma{g}_{ij}=\vec{E}_i \cdot\vec{E}_j$ \citep{Aris1962}. The term $2 \: \ma{A}^i_{\:j} W^j$ on the left-hand side of (\ref{eq_NS4}) may be thought of as a generalization of the Coriolis acceleration. Obviously, this term vanishes if the co-moving reference frame only translates with respect to the laboratory frame, in which case $\ma{R}^{ij}= \delta^{ij}$.\vspace{1mm}

\subsection{General solution of the disturbance problem (\ref{eq_w_adime})}
\label{sec:fourier}
Since (\ref{eq_NS4}) does no longer contain an inhomogeneous advective term, it can be solved  in Fourier space. This yields the general solution of the disturbance problem in the time-dependent basis, $\vec{E}_i$.
In a second step, this solution must be re-expressed  in the Cartesian basis, yielding formally the desired general solution of the disturbance problem (\ref{eq_w_adime}) as
\begin{eqnarray}
\label{dw:fourier}
\vec{\hat{w}}(\vec{k},t) - \vec{\hat{\mathcal{T}}}^{(0)}(\vec{k},t) &= &- \int_0^t\mbox{e}^{ - \int_\tau^t K^2(t-\tau') {\rm\scriptstyle d}\tau'} \:\hat{\ma{G}} \cdot \frac{\mbox{d} \vec{f}^{(0)}(\tau)} {\mbox{d} \tau}   \mbox{d} \tau   \hspace{3cm}  \\
  &&\hspace*{-1.75cm}
  - \int_0^t K^2(t-\tau) \mbox{e}^{ - \int_\tau^t K^2(t-\tau') {\rm\scriptstyle d}\tau'}\Big[ \hat{\ma{G}}  - 
  \ma Y_2(t,\tau)
\cdot \ma{\hat{G}}_2(t-\tau) \cdot \ma{F}(t-\tau)
\Big] \cdot \vec{f}^{(0)}(\tau)  \mbox{d} \tau \,. \nonumber
\end{eqnarray}
Details of the derivation are given in  appendix \ref{AppendixA}. 
Here $\hat{\ma{G}}$ is the Fourier transform of the Green tensor associated with the Stokes equation, namely
\begin{equation}
\hat{\ma{G}} = \frac{1}{k^2} \Big ( \ma{1} - \frac{\vec{k} \vec{k}}{k^2}\Big)\,,
\label{eq:Green}
\end{equation}
such that $\vec{\hat{\mathcal{T}}}^{(0)} = \hat{\ma{G}} \cdot \vec{f}^{(0)}$.
The kernel $K^2$ is directly related to the Cauchy-Green tensor through
\begin{equation}
K^2(\xi) = \vec{k}\cdot \ma{F}(\xi) \cdot \ma{F}^{\sf T} (\xi) \cdot \vec{k} \;,
\label{eq:K2_lab}
\end{equation}
and 
\begin{equation}
\ma{\hat{G}}_2(\xi) = \frac{1}{K^2(\xi)}\Big( \ma{1} - \frac{\ma{F}(\xi) \cdot \ma{F}^{\sf T} (\xi) \cdot \vec{k}\vec{k}   }{K^2(\xi)}\Big)\:,
\label{eq:G2_lab}
\end{equation}
\\
where $\xi=t - \tau$ denotes the time lag. Finally, the second-order tensor
$\ma Y_2(t,\tau)$
 is defined as 
\begin{equation}
 \ma Y_2(t,\tau) = \ma Y(t) \cdot \ma Y^{-1}(\tau)\,,
 \label{y2}
 \end{equation}
  where  $\ma Y(t)$ is the  solution of the fundamental differential equation
\begin{equation}
{\frac{\mbox{d} \ma{Y}(\tau') }{\mbox{d} \tau '} = - 2 \:K^2(t-\tau')\:\ma{\hat{G}}_2(t-\tau')\cdot \ma{A}  \cdot \ma{Y}(\tau')} \quad \mbox{with} \quad 
\ma{Y} (0)  = \ma{1}\:.
\label{eq:fundamental_pb}
\end{equation}
Since (\ref{eq_w_adime}) is linear with respect to time, the solution $\ma Y_2(t,\tau)$ depends only on $\xi$. Note that this property would not hold if the undisturbed flow were time-dependent.
Making use of (\ref{dw:fourier}) in (\ref{eq:force}) and (\ref{eq:torque}), we are now in position to derive a formal expression for the disturbance force and torque, namely
\begin{equation}
\vec{f}'  = -  \ma{M}_1(t) \cdot\vec{u}_s  -  \ma{M}_2(t) \cdot\vec{\omega}_s - \ma{N}_1(t): \ma{S}^\infty  - \epsilon \ma{M}_1(t)  \cdot \int_0^t \ma{K} (t-\tau) \cdot  \tfrac{\rm d}{{\rm \scriptstyle d} \tau}\vec{f}^{(0)} \mbox{d}\tau\:,
\label{eq:force2}
\end{equation} 
and 
\begin{equation}
\vec{\tau}'   =     -  \ma{M}_2^{\sf T}(t) \cdot\vec{u}_s  -  \ma{M}_3(t) \cdot\vec{\omega}_s - \ma{N}_2(t): \ma{S}^\infty  - \epsilon \ma{M}_2^{\sf T}(t) \cdot \int_0^t \ma{K} (t-\tau) \cdot  \tfrac{\rm d}{{\rm \scriptstyle d} \tau}\vec{f}^{(0)} \mbox{d}\tau\:,
\label{eq:torque2}
\end{equation} 
where the kernel $\ma{K}(t)$ may be split in the form
\begin{equation}
\ma{K}(t) = \ma{K}_h(t)  + \int_0^t \ma{K}_i (\xi) \mbox{d}\xi\,,
 \label{eq:K}
\end{equation}
with
\begin{subequations}
\label{eq:khki}
\begin{align}
\ma{K}_h(\xi)& = \frac{1}{8\pi^3}\int\mbox{e}^{ - \int_\tau^t K^2(t-\tau') {\rm\scriptstyle}d\tau'}  \:\hat{\ma{G}} \rm d^3 \vec{k} \,,
\label{eq:Kh}\\
\ma{K}_i(\xi)& =  
\frac{1}{8\pi^3}\int K^2(\xi)   \mbox{e}^{ - \int_\tau^t K^2(t-\tau') {\rm \scriptstyle d}\tau'}\,[\hat{\ma{G}}-
  \ma Y_2(t,\tau)
\cdot \ma{\hat{G}}_2(\xi) \cdot \ma{F}(\xi)]{\rm d}^3 \vec{k} \,. 
 \label{eq:Ki} 
\end{align}
\end{subequations}
Equations (\ref{eq:force2}) and (\ref{eq:torque2}) are the main results of this paper. They provide an explicit expression for the disturbance loads acting on a rigid body with an arbitrary shape moving in a general linear flow in the Saffman limit.
Similar to the  Basset-Boussinesq force acting on a sphere having a time-dependent motion with respect to the fluid \citep[e.g.][]{Landau89}, the instantaneous force acting on a body translating and rotating arbitrarily in a steady linear flow field takes the form of a convolution integral. \\
Remarkably, the kernel $\ma{K}(t)$ defined by (\ref{eq:K}) does not depend on the body shape, the influence of which is entirely accounted for by the resistance tensors $\ma{M}_i$ and $\ma{N}_i$ appearing in the expression for $\vec{f}^{(0)}$ (see \ref{eq_f_0}), and as pre-factors of the convolution integral in (\ref{eq:force2}) and (\ref{eq:torque2}).  That the kernel is independent of the body shape is readily understood by keeping in mind that, at leading order in $\epsilon$, the body is seen as a point force by the far-field flow.  Thanks to this crucial property, the kernel may be determined once and for all and the method can then be applied to any body shape, provided that the resistance tensors are known.  
It is important to note that this state of affairs drastically differs from the problem of the leading-order inertial corrections to the rotational dynamics of neutrally-buoyant non-spherical bodies immersed in shear flows. Such corrections were first derived by \cite{subramanian2005} for rodlike bodies, then by \cite{einarsson2015a} and \cite{Dabade2016} for spheroids with arbitrary aspect ratios. In all cases, the corrections were obtained using a regular perturbation expansion in powers of $\epsilon$ in which the first nonzero correction to the torque was found to occur at ${\mathcal{O}}(\epsilon^2)$. That this correction is provided by a regular expansion indicates that it is driven by the near-field flow, and thus depends directly upon the body shape, in contrast to the ${\mathcal{O}}(\epsilon)$-correction derived here. This suggests that, regarding the translational dynamics, higher-order corrections to present results, similar to the so-called second-order Saffman's lift force \citep{Saffman65,McLaughlin91}, may incur direct dependencies on the body shape as well.\\
As a last point, we stress that results (\ref{eq:force2})-(\ref{eq:Ki}) are distinct from the expression of the unsteady force and torque acting on an arbitrarily-shaped body derived by \cite{Gavze90}. His work is concerned with an entirely different limit of the problem, where unsteady contributions are of the same order as the quasi-steady Stokes drag (this limit is obtained by setting $\mbox{Re}_s\rightarrow0$, $\mbox{Re}_p\rightarrow0$ and $\mbox{Re}_s\mbox{Sl}={\mathcal{O}}(1)$ in (\ref{eq_w_1_2})). Under such conditions, the problem is equivalent to the time-dependent Stokes equation solved by \cite{Boussinesq85} and \cite{Basset88} past a sphere. However, when generalized to bodies of arbitrary shape, the unsteady part of the solution (which yields added-mass and `history' effects) is found to involve two supplementary shape-dependent tensors similar to the `grand-resistance' tensor in (\ref{eq_wrench}). Here in contrast, as inertial effects are assumed to provide only small corrections to the quasi-steady Stokes drag, they are connected to the body shape exactly in the same way as the primary drag, namely through the resistance tensors $\ma{M}_1$ and $\ma{M}_2$. 
\section{The kernel in the two canonical bidirectional linear flows}
\label{bidir}
 To prove the versatility of the approach derived in the previous section, we first specialize it to the canonical cases of a solid-body rotation flow and a planar elongational flow, respectively. In the former case, transforming the disturbance flow equations in the co-moving coordinate system merely corresponds to performing a change of reference frame, which makes this situation a compulsory test case.
  \subsection{Solid-body rotation }
As reviewed in the introduction, this configuration is well documented in the literature. Interestingly, determining the inertial drag correction experienced by a sphere translating with a constant velocity along the axis of a solid-body rotation flow was the question that motivated \cite{Childress64} to design the MAE approach in the way that later became standard in the class of problems considered here; his predictions were checked experimentally by \cite{Maxworthy65}. \vspace{2mm}\\
\noindent In the laboratory frame, the base flow reads
\begin{equation}
\vec{U}^\infty(\vec{x} ) = \ma{A} \cdot \vec{x}\,, \quad \mbox{with} \quad
\ma{A} = \left( \begin{array}{ccc} 
0 &-1 & 0\\
1&0&0\\
0& 0&0
\end{array} \right)\:.
\end{equation}
The matrix exponentiation (\ref{eq:Exp_A}) provides the deformation gradient tensor $\ma{F}$ in the form
\begin{equation}
\ma{F}(t) = \left( \begin{array}{ccc} 
\cos\,t &-\sin\,t & 0\\
\sin\,t&\cos\,t&0\\
0& 0&1
\end{array} \right) \:.
\label{FR}
\end{equation}
Using (\ref{FR}), (\ref{eq:K2_lab}) yields $K^2(\xi) = k^2$, hence $\int_\tau^t K^2(t-\tau') \mbox{d}\tau' =k^2\xi$, 
so that the fundamental problem (\ref{eq:fundamental_pb}) reduces to
\begin{equation}
\left\{ \begin{array}{l} 
\displaystyle{\frac{\mbox{d} \ma{Y}(\tau') }{\mbox{d} \tau '} =  \frac{2}{k^2}\left( \begin{array}{ccc} 
 k_1 k_2 &  (k_2^2+k_3^2) & 0 \\
-(k_1^2+k_3^2) & - k_1 k_2 & 0 \\
 k_2 k_3 & - k_1 k_3 & 0
\end{array} \right)  \cdot \ma{Y}(\tau')} \\
\\
\ma{Y} (0)  = \ma{1}\:.
\end{array}
\right.
\label{eq:fundamental_pb_rotation}
\end{equation}
As the right-hand side of (\ref{eq:fundamental_pb_rotation}) depends upon the time lag only, integration can be achieved again through a matrix exponentiation. Setting $Z=\frac{2 k_3 \xi}{k}$, this leads to the solution
\begin{equation}
\ma{Y}_2(\xi) = \left( \begin{array}{ccc} 
\frac{k_1k_2}{k_3 k} \sin Z +\cos Z & \frac{k_2^2 +k_3^2}{k_3 k} \sin Z & 0 \\
-\frac{k_1^2 +k_3^2}{k_3 k} \sin Z  & -\frac{k_1k_2}{k_3 k} \sin Z +\cos Z  & 0 \\
\frac{k_2}{k} \sin Z-\frac{k_1}{k_3} \cos Z + \frac{k_1}{k_3} & -\frac{k_1}{k} \sin Z -\frac{k_2}{k_3} \cos Z + \frac{k_2}{k_3}  & 1
\end{array} \right) \:.
\end{equation}
Then the two kernels $\ma{K}_h$ and $\ma{K}_i$ involved in (\ref{eq:Kh}) and (\ref{eq:Ki}) are obtained in the form
\begin{equation}
6\pi \ma{K}_h(\xi) = \frac{\ma{1}}{\sqrt{\pi \xi}}  \:,
\quad \mbox{} \quad  
6 \pi \ma{K}_i(\xi) = \left( \begin{array}{ccc} 
I_1(\xi)  & I_2(\xi ) & 0 \\
-I_2(\xi)  & I_1(\xi)  & 0 \\
0 & 0& I_3(\xi) 
\end{array} \right) \,,
\label{Krot}
\end{equation}
where 
\begin{eqnarray}
\label{I1}
I_1(\xi) &= &\frac{1}{16} \frac{3 \sin\xi\cos^2\xi -6 \xi^2\sin\xi -3 \xi \cos\xi+8 \xi^3}{\sqrt{\pi} \xi^{9/2}}\,, \\
\label{I2}
I_2(\xi) &= &-\frac{3}{16} \frac{\cos\xi-\cos^3\xi-2 \xi^2\cos\xi+\xi\sin\xi}{ \sqrt{\pi} \xi^{9/2}}\,, \\
\label{I3}
I_3 (\xi) &= &-\frac{1}{8} \frac{-4 \xi^3+3 \sin\xi \cos\xi-6 \xi \cos^2\xi+3 \xi}{\sqrt{\pi} \xi^{9/2}}\:. 
\end{eqnarray}

\noindent After integrating by parts, these kernels are found to be identical to those obtained by \cite{Candelier08}. Expanding (\ref{Krot}) for short times and setting $\tau=0$ yields $\ma K(t)$ in the form\\
\begin{equation}
6\pi \ma K(t)
= \frac{1}{\sqrt{\pi}}
\left(\begin{array}{ccc}
 t^{-1/2}  + \tfrac{1}{10}  {t^{3/2}} & - \tfrac{1}{75} t^{5/2} & 0 \\
         \tfrac{1}{75} t^{5/2}       & t^{-1/2}  + \tfrac{1}{10}  {t^{3/2}} & 0 \\
0 & 0 & t^{-1/2}  + \frac{2}{15}  {t^{3/2}}
\end{array}
\right) +  \ldots
\label{Krotshort}
\end{equation}
The contribution to the hydrodynamic force associated with the $t^{-1/2}$ diagonal terms (which result from the kernel $\ma{K}_h$ in (\ref{Krot})) corresponds the usual Basset-Boussinesq `history' force \citep{Landau89}. Inertial corrections due to the background linear flow result from the kernel $\ma{K}_h$. They are seen to grow as $t^{3/2}$ on the diagonal, faster than off-diagonal corrections corresponding to a lift force, which grow as $t^{5/2}$. Integrating $\ma{K}_i $ over time, the quasi-steady kernel corresponding to the long-time limit $t\rightarrow\infty$ is found to be
\begin{equation}
6\pi \overline{\ma K} = 6\pi   \int_0^\infty \ma{K}_i (\xi) \mbox{d}\xi =
\left( \begin{array}{ccc} 
\frac{3 \sqrt{2} (19+9 \sqrt{3})}{280}  & -\frac{3 \sqrt{2} (19-9 \sqrt{3})}{280} & 0 \\
\frac{3 \sqrt{2} (19-9 \sqrt{3})}{280} & \frac{3 \sqrt{2} (19+9 \sqrt{3})}{280} & 0 \\
0 & 0& \frac{4}{7}
\end{array} \right)\,.
\end{equation}
This is the result obtained independently by \cite{Gotoh90} and \cite{Miyazaki95b} through the MAE and IF approaches, respectively. The ${[\overline{\ma K}]^3}_3$ component was determined much earlier by \cite{Childress64}. Inertial effects due to the solid-body rotation are seen to increase the drag whatever the direction of the slip velocity, with a slightly larger pre-factor when the body moves along the rotation axis ($6\pi{[\overline{\ma K}]^3}_3\approx0.571$) than within the plane of the flow ($6\pi{[\overline{\ma K}]^1}_1=6\pi{[\overline{\ma K}]^2}_2\approx0.542$). Inertial effects also induce a small nonzero lift component ($6\pi{[\overline{\ma K}]^2}_1=-6\pi{[\overline{\ma K}]^1}_2\approx0.052$) which is centrifugal if the sphere is at rest in the laboratory frame. More generally, this lift component is centrifugal (resp. centripetal)  if the sphere translates in such a way that it lags behind (resp. leads) the fluid. Note that the same situation was considered by \cite{Drew78}; however his calculation erroneously predicted the lift component to be zero.\vspace{2mm}\\
\begin{figure} 
\includegraphics[width=190pt]{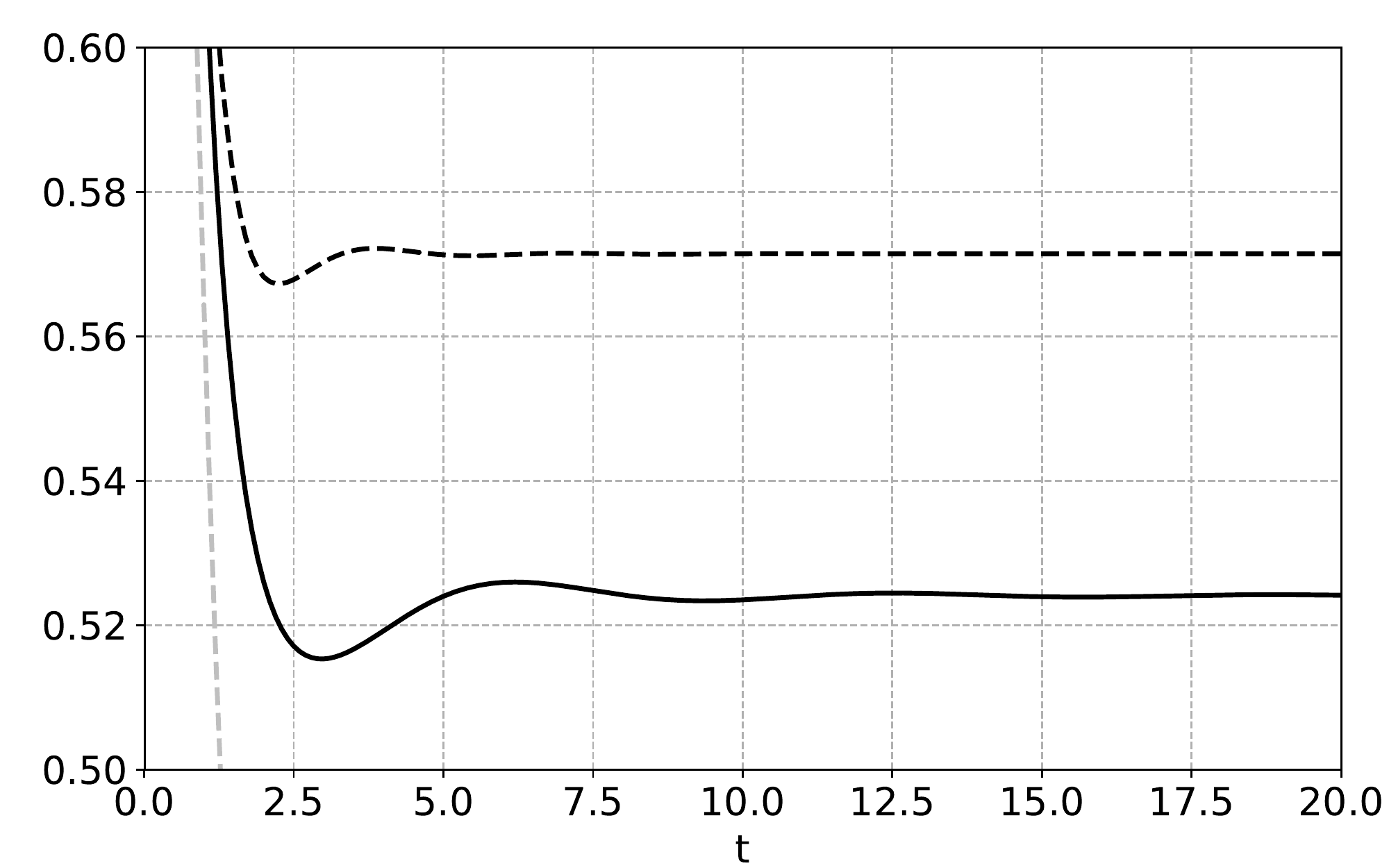}
\includegraphics[width=190pt]{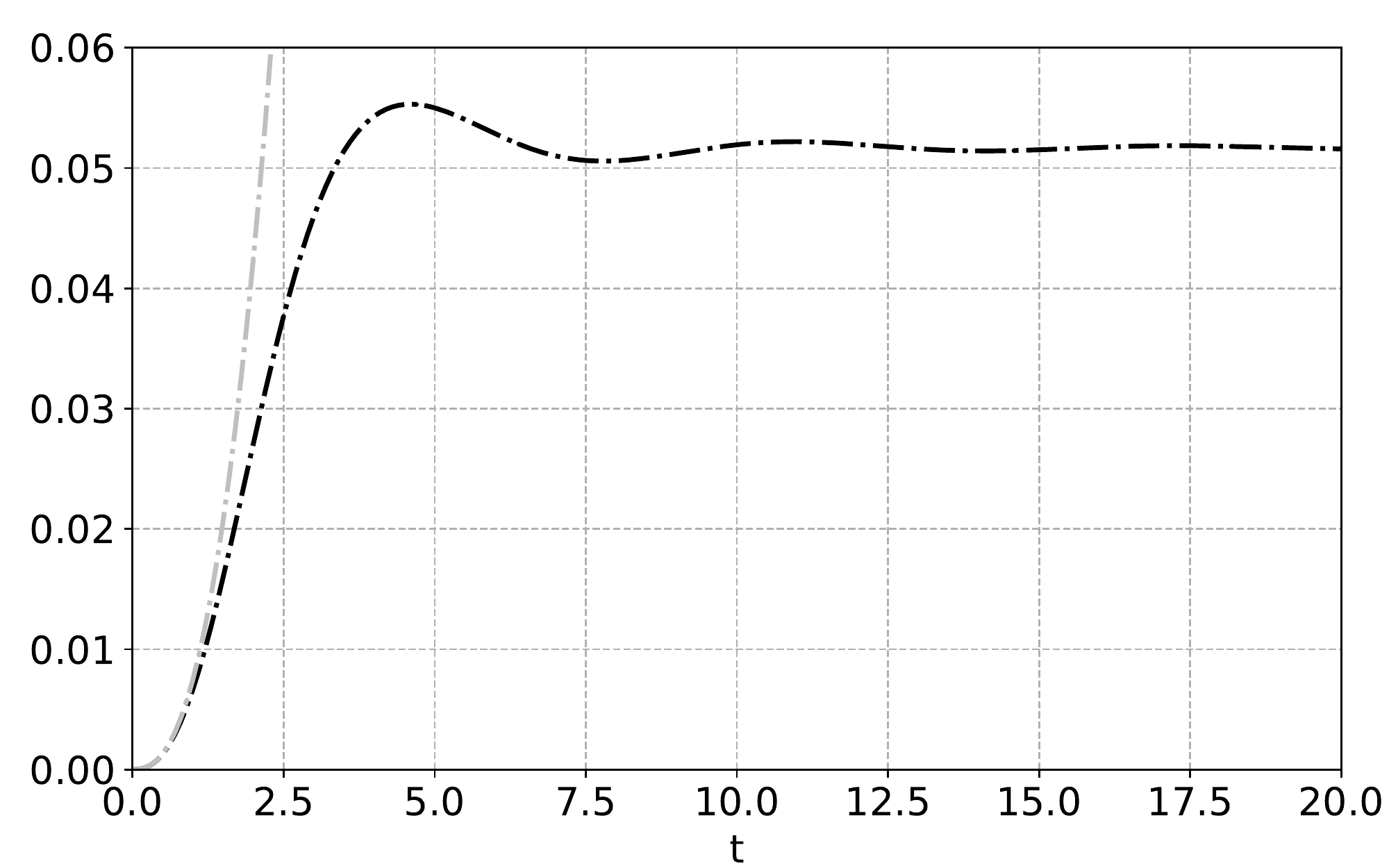}
\caption{Time variation of the kernel $\ma{K}$ in a solid-body rotation flow. Solid line: $6\pi {[\ma{K}]^1}_1=6\pi {[\ma{K}]^2}_2$; dashed line: $ 6 \pi {[\ma{K}]^3}_3$; grey dashed line: $t^{-1/2}$ short-time behaviour resulting from the contribution ${\ma{K}}_h(t)$ in (\ref{Krot}); black dash-dotted line: $ 6 \pi { |[\ma{K}]^1}_2$; grey dash-dotted line: short-time expansion $ 6 \pi {[\ma{K}]^1}_2 \sim \frac{1}{75\sqrt\pi} t^{5/2}$.  }
\label{Kernel_rot}
\end{figure}
Figure \ref{Kernel_rot} shows how $\ma{K}(t)$ reaches the above steady state; according to (\ref{eq:force2}), the inertial corrections to the force directly follow this evolution if the slip velocity is set abruptly to a nonzero constant value at time $t=0$. The diagonal components are seen to reach levels close to their steady-state value in approximately two time units. In contrast, it takes approximately twice as long for the off-diagonal component, ${\ma{K}^1}_2$, to reach a quasi-converged level. This is due to the different growth rates identified in (\ref{Krotshort}). In all cases, damped oscillations with a period $T_o=2\pi$ corresponding to the imposed rotation rate $\frac{1}{2}|\vec{\Omega}^{\infty}|=1$ take place subsequently.

\subsection{Planar elongational flow}
\label{elongflow}
As a second example, we consider the purely extensional planar flow defined by
\begin{equation}
\vec{U}^\infty(\vec{x}) = \ma{A} \cdot \vec{x}\,, \quad \mbox{with} \quad
\ma{A} = \left( \begin{array}{ccc} 
1 &0 & 0\\
0&-1&0\\
0& 0&0
\end{array} \right)\:.
\label{ecoulelong}
\end{equation}
This configuration was first considered by \cite{Drew78} in the steady-state limit, using the MAE approach, then by \cite{PerezMadrid90} using the IF method. However, as discussed in section \ref{intro}, the results of P\'erez-Madrid \textit{et al.} are not correct, since they neglected the inhomogeneous term arising from (\ref{inhom}). The same warning applies to the conclusions of \cite{Bedeaux87} who also examined the corresponding time-dependent situation.\vspace{2mm}\\
\noindent The deformation-gradient tensor corresponding to (\ref{ecoulelong}) is
\begin{equation}
\ma{F}(t) = \left( \begin{array}{ccc} 
\mbox{e}^t &0 & 0\\
0&\mbox{e}^{-t}&0\\
0& 0&1
\end{array} \right) \:.
\end{equation}
It follows from (\ref{eq:K2_lab}) that
\begin{equation}
K^2(\xi) = k_1^2\mbox{e}^{2\xi}+k_2^2 \mbox{e}^{-2\xi}+k_3^2 \,,
\end{equation}
and thus
\begin{equation}
\int_\tau^t K^2(t-\tau') \mbox{d}\tau' =  \frac{1}{2} k_1^2( \mbox{e}^{2\xi}-1)+\frac{1}{2} k_2^2(1-\mbox{e}^{-2\xi}) +k_3^2\xi\:.
\label{K2elong}
\end{equation}
Equations (\ref{eq:G2_lab})-(\ref{eq:fundamental_pb}) show that the fundamental problem to solve is 
\begin{equation}
\left\{ \begin{array}{l} 
\displaystyle{\frac{\mbox{d} \ma{Y}(\tau') }{\mbox{d} \tau '} = \frac{2}{k_1^2 \mbox{e}^{2(t-\tau')}+k_2^2\mbox{e}^{-2(t-\tau')}+k_3^2} \left( \begin{array}{ccc} 
& &\\
-\left(k_2^2 \mbox{e}^{-2(t-\tau')}+k_3^2\right) &-\mbox{e}^{2(t-\tau')}k_1 k_2 & 0\\
&& \\
\mbox{e}^{-2(t-\tau')} k_1 k_2 & \left(k_1^2 \mbox{e}^{2(t-\tau')}+k_3^2\right)&  0\\
k_1 k_3 & -k_2 k_3 & 0
\end{array} \right)  \cdot \ma{Y}(\tau')}\,, \\
\\
\ma{Y} (0)  = \ma{1}\:.
\end{array}
\right.
\label{eq:fundamental_pb_Elongational_2D}
\end{equation}

This differential equation has the solution 
\begin{equation}
\ma{Y}_2(\xi) = \left( \begin{array}{ccc} 
\displaystyle{\frac{k_1^2 +\mbox{e}^{-2\xi} (k_2^2+k_3^2)}{k^2}}& \displaystyle{\frac{k_1k_2 (1-\mbox{e}^{2\xi})}{k^2} }& 0 \\
& & \\
\displaystyle{\frac{k_1k_2 (1-\mbox{e}^{-2\xi})}{k^2}} & \displaystyle{\frac{k_2^2 +\mbox{e}^{2\xi} (k_1^2+k_3^2)}{k^2}}& 0\\
& & \\
\displaystyle{\frac{k_1k_3 (1-\mbox{e}^{-2\xi})}{k^2} } & \displaystyle{\frac{k_2k_3 (1-\mbox{e}^{2\xi})}{k^2} } & 1
\end{array} \right)\,. 
\label{solelong}
\end{equation}

\noindent The kernel $\ma{K}$ is readily obtained after inserting (\ref{K2elong}) and (\ref{solelong}) into (\ref{eq:Kh}) and (\ref{eq:Ki}) and performing integrations. At short times, the non-zero components of $\ma{K}$ may be obtained in the form of a regular expansion in $t$. Truncating this expansion to ${\mathcal{O}}(t^{3/2})$ terms, we find  
\begin{equation}
6\pi \ma{K}(t) = \frac{1}{\sqrt{\pi}}\left( \begin{array}{ccc}
t^{-1/2}+\tfrac{7}{10}t^{1/2} - \tfrac{1}{105}t^{3/2}&   0 & 0 \\
0& t^{-1/2} - \tfrac{7}{10} t^{1/2}  - \tfrac{1}{105}t^{3/2}&    0 \\
0& 0& t^{-1/2} + \tfrac{11}{210} t^{3/2}\\
\end{array}\right)+...
\label{kernelelong}
\end{equation}

\begin{figure} 
\begin{center}
\begin{psfrags}
\psfrag{t}[c][c][1]{$t$}
\includegraphics[width=0.9\textwidth]{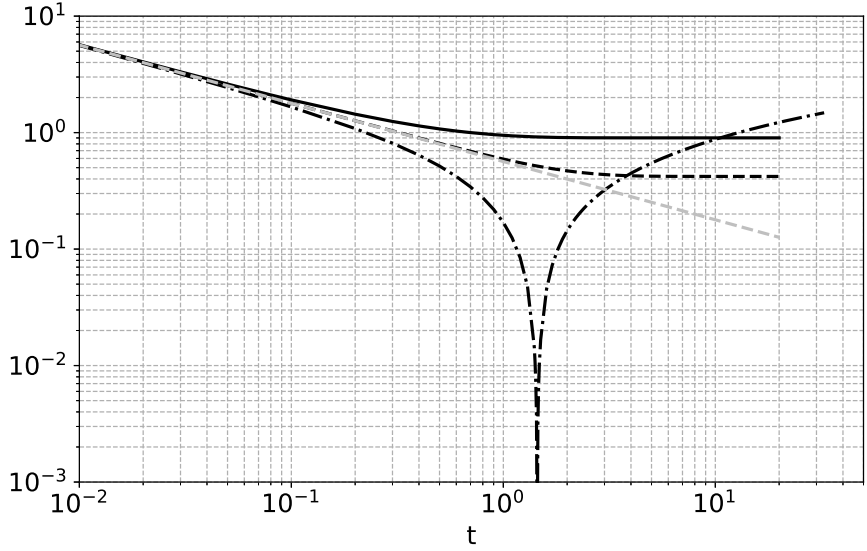}
\end{psfrags}
\end{center}
\caption{Time variation of the kernel $\ma{K}$ in a planar elongational flow. Solid line: $6\pi {[\ma{K}]^1}_1$; dashed line: $ 6 \pi {[\ma{K}]^3}_3$; dash-dotted line: $ 6 \pi { |[\ma{K}]^2}_2 |$; grey dashed line: $t^{-1/2}$ short-time behaviour. The component ${[\ma{K}]^2}_2$ switches from positive to negative at $t\approx1.4$.  }
\label{Kernel}
\end{figure}

\noindent Again, the leading-order term in this expansion evolves as $t^{-1/2}$, a behaviour characterizing the response of the hydrodynamic force to an impulsive velocity change \citep{Landau89}. The two $t^{1/2}$-contributions in (\ref{kernelelong}) have the same magnitude and are in agreement with the high-frequency behaviour determined by \cite{Bedeaux87}.\vspace{2mm}\\
\noindent To determine the steady-state limit of $\ma K$, we evaluated the $\vec k$-integrals involved in (\ref{eq:Kh}) and (\ref{eq:Ki}) numerically. 
 The result is shown in figure \ref{Kernel}. After the three components separate for $t\approx0.1$, ${[\ma{K}]^1}_1$ and ${[\ma{K}]^3}_3$ gently reach their steady state value for $t={\mathcal{O}}(1)$. In contrast, the component ${[\ma{K}]^2}_2(t)$ corresponding to the compressional direction of the strain sharply decreases and becomes negative for $t\approx1.4$. At longer times, its absolute value increases and becomes of ${\mathcal{O}}(1)$. So far, despite various attempts to stretch the integrand in the vicinity of $\vec k=\vec 0$ (which yields the dominant contribution to the steady-state components of $\ma K$), we have been unable to compute  ${[\ma{K}]^2}_2(t)$ beyond $t \approx 32$, where we find (using Maple\textsuperscript{\textregistered}) $  6\pi{[\ma{K}]^2}_2\approx-1.48$. At the present stage, considering that the tendency for the absolute value of ${[\ma{K}]^2}_2$ to increase goes on at later times, we find the steady-state kernel to be
\begin{equation}
6\pi \overline{\ma K} =6\pi   \int_0^\infty \ma{K}_i (\xi) \mbox{d}\xi \simeq \left( \begin{array}{ccc}
0.901 &   0 & 0 \\
0&  {6\pi[\ma{\overline{K}}]^2}_2<-1.48 &    0 \\
0& 0& 0.420\\
\end{array}\right)\,.
\label{kernelelongst}
\end{equation}
Only the diagonal components are nonzero in (\ref{kernelelong}) and (\ref{kernelelongst}), and they all differ in magnitude. Consequently, if the body does not move along one of the principal directions of the strain, it experiences a transverse or lift force. For instance, suppose a sphere moves ahead of the fluid with a unit slip velocity along the first bisector of the $(\vec e_1,\,\vec e_2)$ plane. Then, according to (\ref{kernelelong}), it experiences a growing transverse force $\vec f_T(t)=-18\pi^2([\ma{K}]^1_1-[\ma{K}]^2_2)(\vec e_1-\vec e_2)=\frac{21}{10}(2\pi)^{1/2}\epsilon t^{1/2}(\vec e_2-\vec e_1)$ at short times, which eventually becomes $\vec {\overline{f}}_T=3\pi(0.901-6\pi{[\ma{\overline{K}}]^2}_2)\epsilon\frac{(\vec e_2-\vec e_1)}{\sqrt{2}}$ in the steady-state limit. As ${[\ma{\overline{K}}]^2}_2$ is expected to be negative, this transverse force tends to deviate the sphere toward the compressional $\vec e_2$-axis at both short and long times. This is qualitatively consistent with the conclusion of \cite{Drew78} who, in present notations, found $\vec {\overline{f}}_T=3.012\pi\epsilon\frac{(\vec e_2-\vec e_1)}{\sqrt{2}}$. However this pre-factor is uncertain because Drew's result for the kernel component corresponding to the $\vec e_1$-extensional direction is $6\pi{[\ma{\overline{K}}]^1}_1=0.602$, instead of $0.901$ in (\ref{kernelelongst}).

\section{The kernel in a linear shear flow}
\label{sec:kernel}
\begin{figure}
\begin{center}
\begin{psfrags}
\psfrag{t}[c][c][1]{$t$} 
\includegraphics[width=\linewidth]{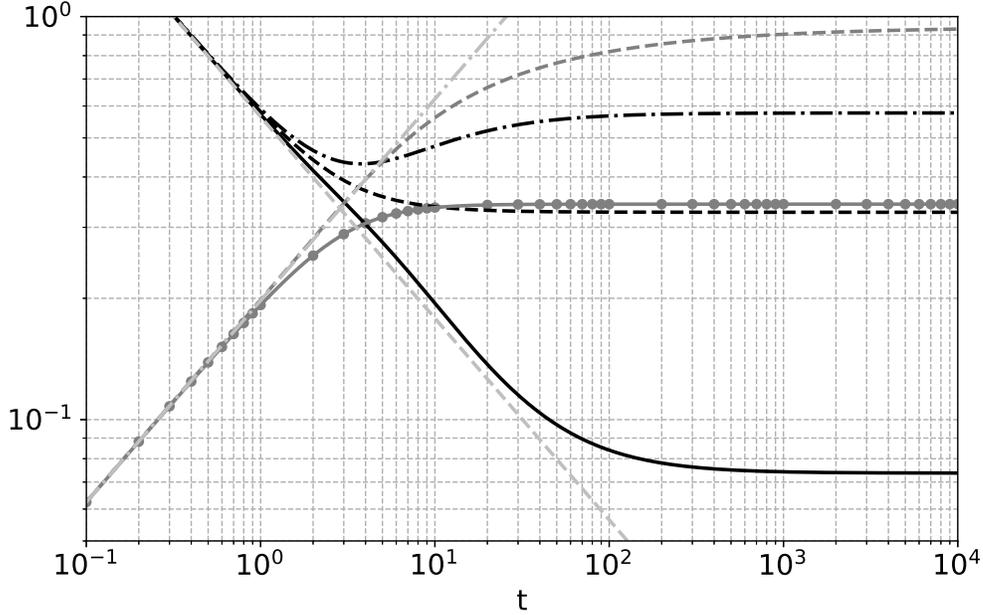}
\end{psfrags}
\caption{Time variation of the kernel $\ma{K}$ in a linear shear flow. Black lines correspond to the diagonal components (i.e. the inertial corrections to the drag force), with $6\pi{[\ma{K}]^1}_1$
(solid line), $6\pi{[\ma{K}]^2}_2$ (dash-dotted line), and  $6\pi{[\ma{K}]^3}_3$ (dashed line). Dark grey lines correspond to the off-diagonal components, 
 with $6 \pi{[\ma{K}]^1}_3$ (dashed line), and $6\pi{[\ma{K}]^3}_1$ (solid line); the latter is the time-dependent counterpart of the Saffman lift force. Circles correspond to the inverse Fourier transform of the results obtained in the frequency domain by \cite{Asmolov99}. Pale grey lines correspond to the $t^{-1/2}$-Basset-Boussinesq kernel (dashed line), and to the off-diagonal components of the kernel derived by \cite{Miyazaki95a} in the short time limit (dash-dotted line).
 } \label{comp_k}
\end{center}
\end{figure}

\noindent We now consider the more widely studied case of a linear shear flow in which $\ma{A}$ takes the form
\begin{equation}
\ma{A} = \left( \begin{array}{ccc} 
0 &0 & 1\\
0&0&0 \\
0& 0&0\\
\end{array} \right)\:,
\label{eq:flow_shear}
\end{equation}
\\
In that case, the unit vector $\vec{e}_1$ points in the flow direction,  $\vec{e}_3$ points in the shear direction, and $\vec{e}_2$ is aligned with the direction of the undisturbed vorticity, such that $\nabla\times\vec{U}^\infty=-\vec{e}_2$. Exponentiating $\ma{A}$, (\ref{eq:Exp_A}) implies
\begin{equation}
\ma{F}(t) = \left( \begin{array}{ccc} 
1 &0 & t\\
0&1&0\\
0& 0&1
\end{array} \right) \:.
\end{equation}
Inserting this result into (\ref{eq:K2_lab}) yields $K^2(\xi) = k^2 +2 k_1 k_3 \xi+ k_1^2 \xi^2 $, so that
\begin{equation}
\int_\tau^t K^2(t-\tau') \mbox{d}\tau' =  k^2 \xi + k_1 k_3 \xi^2 +\frac{1}{3} k_1^2 \xi^3 \,.
\end{equation}
Then (\ref{eq:fundamental_pb}) takes the form
\begin{equation}
{\frac{\mbox{d} \ma{Y}(\tau') }{\mbox{d} \tau '} = \frac{2}{k^2 +2 k_1 k_3 (t-\tau')+ k_1^2 (t-\tau')^2} \Bigg( \begin{array}{ccc} 
\scriptstyle 0 & \scriptstyle 0 &\scriptstyle  -  [k_2^2+k_3^2+k_3 k_1 (t-\tau')]\\
\scriptstyle 0 &\scriptstyle 0 &\scriptstyle  k_1 k_2\\
\scriptstyle 0 &\scriptstyle 0 &\scriptstyle [k_1k_3+k_1^2 (t-\tau')]\end{array}\Bigg)
 \cdot \ma{Y}(\tau')} \,,
\label{eq:fundamental_pb_Saffman}
\end{equation}
with the initial condition $\ma{Y} (0)  = \ma{1}$. 
The solution of (\ref{eq:fundamental_pb_Saffman}) yields the second-order tensor $\ma Y_2$ defined in (\ref{y2}) in the form
\begin{equation}\begin{split}
\ma Y_2(\xi)&=
\left( \begin{array}{ccc} 
1& 0 & \displaystyle{- \tfrac{(k^2+2 k_3 k_1 \xi+k_1^2 \xi^2) k_2^2 S(k_1,\:k_2,\:k_3,\:\xi)}{(k1^2+k2^2) \sqrt{k_1^2 (k_1^2+k_2^2)}} -\tfrac{\xi(k_3k_1^3 \xi+k_2^2 k^2+2 k_1^2 k_3^2)}{(k_1^2+k_2^2) k^2} }\\
\\
0 & 1& \displaystyle{\tfrac{(k^2+2 k_3 k_1 \xi+k_1^2 \xi^2) k_2k_1S(k_1,\:k_2,\:k_3,\:\xi)}{(k1^2+k2^2) \sqrt{k_1^2 (k_1^2+k_2^2)}}  +  \tfrac{ \xi (k_1^2+k_2^2-k_3^2-k_3 k_1 \xi) k_2 k_1}{(k_1^2+k_2^2) k^2}} \\
\\
0 & 0 & \displaystyle{\tfrac{k^2+2 k_3 k_1 \xi+k_1^2 \xi^2}{k^2}}
\end{array} \right) \,,
\end{split}
\end{equation}
where the function $S(k_1,\:k_2,\:k_3,\:\xi)$ is given by
\begin{equation}
S(k_1,\:k_2,\:k_3,\:\xi) = \tan^{-1}\!\Big(\frac{k_1(k_3+k_1 \xi)}{ \sqrt{k_1^2 (k_1^2+k_2^2)}}\Big)-\tan^{-1}\!\Big(\frac{k_1k_3}{ \sqrt{k_1^2 (k_1^2+k_2^2)}}\Big)\:.
\end{equation}
The kernel $\ma K(t)$ is obtained after inserting the above expressions into (\ref{eq:K})-(\ref{eq:khki}) and performing the required integrations. Figure \ref{comp_k} shows
the corresponding result for each nonzero component of $\ma K(t)$, integrations having again been performed with Maple\textsuperscript{\textregistered}.
At short times, each component exhibits a power law form, albeit with a different exponent for the diagonal and off-diagonal components.
Also shown are numerical data (circles) for the ${[\ma{K}]^3}_1$ component which corresponds to the Saffman lift force in the limit $t\rightarrow\infty$ in the case of a sphere, i.e. with $\vec{f}^{(0)}=6\pi\vec{e}_1$. To obtain these data, we numerically performed the inverse Fourier transform of the frequency-dependent expression derived by \cite{Asmolov99}, as described by \cite{Candelier07}. \cite{Asmolov99} could compute the full frequency dependency for this specific kernel component because in that specific case, the partial differential equation (\ref{eq_w_hat}) simplifies to an ordinary differential equation.
We did not succeed in simplifying the kernel components (\ref{eq:Kh}) and (\ref{eq:Ki}) for arbitrary values of $t$. This is why we illustrate the physical mechanisms at play and compare present findings with available results \citep{Harper68,Miyazaki95a} in the two limit cases $t\ll1 $ and $t\gg 1$. 

\subsection{Short-time limit}
\label{subsec:limit_small_t} 

In the former limit, similar to the case of the purely elongational flow, the non-zero components of $\ma{K}$ may be obtained in the form of a regular expansion with respect to $t$. Keeping only the first two terms in each infinite series, $\ma{K}$ reduces to \vspace{2mm}
\begin{equation}
6\pi \ma K(t)
= \frac{1}{\sqrt{\pi}}
\left(\begin{array}{ccc}
 t^{-1/2}  + \tfrac{1}{70}  {t^{3/2}} & 0 & \frac{7}{20}t^{1/2} + \frac{1}{800}  {t^{5/2}}\\
                                  0                                 & t^{-1/2}  + \frac{13}{280}  {t^{3/2}} &0 \\
\frac{7}{20}t^{1/2} - \frac{53}{5600}  {t^{5/2}} &0 & t^{-1/2}  + \frac{13}{420}  {t^{3/2}}
\end{array}
\right) + \ldots 
\label{shortt}
\end{equation}
\\
Not surprisingly, the leading-order behaviour of the diagonal terms is again found to behave as $t^{-1/2}$. As is well known, the corresponding contributions, which yield the classical Basset-Boussinesq `history' force, result from vorticity diffusion across the boundary layer that develops around the body after the flow is abruptly started. These effects only involve the inner solution corresponding to $|\vec{r}|\ll\epsilon^{-1}$ as discussed in \S\ref{asympt}, since vorticity stays concentrated in the neighbourhood of the body for $t\ll1$. The initial $t^{-1/2}$ decrease of the drag components may then be readily understood by equating the rate of work of the drag force to the viscous dissipation throughout the fluid. Since the boundary layer thickness grows as $t^{1/2}$, velocity gradients within it decay as $t^{-1/2}$, making the local dissipation rate decrease as $t^{-1}$. Therefore the integral of this dissipation throughout the boundary layer volume decreases as $t^{-1/2}$ and the drag force changes accordingly. \\
\noindent Figure \ref{comp_k} shows that the diagonal components of $\ma{K}$ (black lines) start to depart from the $t^{-1/2}$ behaviour after a few time units, which typically 
corresponds to the time it takes for the vorticity to reach the Saffman distance, $\ell_s=\epsilon^{-1}$. 
The next-order terms are the signature of inertial effects resulting from the increasing role of vorticity advection at distances from the body larger than $\ell_s$. The $t^{3/2}$ terms involved in the diagonal components differ from one component to the other, due to the anisotropy of the base flow. \vspace{2mm}\\
Let us now consider the off-diagonal components in (\ref{shortt}), depicted by dark grey lines in figure \ref{comp_k}. Only those corresponding to a slip velocity lying in the plane of the shear are nonzero. The component ${[\ma{K}]^3}_1$ corresponds to a force directed along the shear (hence at right angle from the streamlines) when the slip velocity is aligned with the undisturbed flow. This is the component that yields the Saffman lift force in the long-term limit. The component ${[\ma{K}]^3}_1$ corresponds to a force aligned with the streamlines when the slip velocity lies in the direction of the shear; it was first computed by \cite{Harper68}.
Both lift components cannot exist in the creeping flow limit, owing to reversibility \citep{Bretherton1962}, and are therefore due to fluid inertia effects. As already observed in the elongational flow, these two off-diagonal components are equal in the short-time limit. They are identical to the leading-order inertial corrections of the diagonal components in (\ref{kernelelong}) and agree with the high-frequency asymptote of the mobility tensor, $\ma{U}(\omega)=(\ma{M}_1)^{-1}(\omega)$, computed by \cite{Miyazaki95a}, which in dimensional variables writes in every linear flow (equation (5.10) in their paper) 
\begin{equation}
\label{eq:mu}
\ma{U}(\omega) \sim -\left(\frac{-\mathtt{i} \omega a^2}{\nu}\right)^{1/2} \left(\ma{1}+\frac{7}{40} \frac{\ma{A} + \ma{A}^{\sf T}}{\mathtt{i} \omega}\right)\:,
\end{equation}
where $\omega$ denotes the radian frequency. 
The short-time contribution in ${[\ma{K}]^3}_1$ is also identical to that determined by \cite{Asmolov99}. \\
\noindent The $t^{1/2}$ short-time evolution of these off-diagonal components may be understood by considering how nonlinear effects modify the vorticity disturbance $\boldsymbol\omega=\nabla\times\vec{w}$, especially how they tilt the upstream vorticity $\nabla\times\vec{U}^\infty$ oriented along the $\vec{e}_2$-direction to generate a nonzero vorticity component oriented along the $\vec{e}_1$-direction in the wake. This streamwise vorticity component is known to be the key ingredient  yielding a nonzero lift force on a three-dimensional body \citep{Lighthill56}. According to (\ref{eq_w_1_3}), the vortex stretching/tilting term writes $ \mbox{Re}_s\{\ma{A}\cdot\boldsymbol\omega+(\nabla\times(\ma{A}\cdot\vec{r}))\cdot\nabla\vec{w}\}$, so that its streamwise component is $\mbox{Re}_s\{\omega_3- (\vec{e}_2\cdot\nabla) w_1\}$. At short time, $\omega_3$ and $(\vec{e}_2\cdot\nabla) w_1$ decay as $t^{-1/2}$ within the boundary layer, owing to the $t^{1/2}$ thickening of the latter, and so does the vortex stretching/tilting term. To balance this decay, the time rate-of-change of $\omega_1$ (and the diffusion term $\nabla^2\omega_1$) must decay at the same rate, which results is a $t^{1/2}$-growth of $\omega_1$. Since $\omega_1=(\vec{e}_2\cdot\nabla) w_3-(\vec{e}_3\cdot\nabla) w_2$, the growth of $\omega_1$ induces inertial corrections to the transverse velocity components $w_2$ and $w_3$ that also grow as $t^{1/2}$. This in turn results in a similar growth of the transverse pressure gradient, which yields the observed $t^{1/2}$-growth of the non-diagonal components of the force on the body. That these two components are identical for $t\ll1$ may readily be understood by considering successively a sphere translating along the $\vec{e}_1$- and $\vec{e}_3$-directions, with the same slip velocity. The corresponding two inner solutions are identical (up to a switch in the dependency with respect to the $x_1$ and $x_3$ coordinates), and so is the vorticity distribution about the sphere. Hence the initial stretching and tilting of the vorticity in the wake have the same magnitude in both configurations, an so do the two components of the lift force. 
Figure \ref{comp_k} indicates that these two components separate beyond a time of the order of a few units, as already observed for the diagonal terms. The reason for this separation is discussed in the next paragraph.

\subsection{Long-time limit}
\label{subsec:long_t} 
To determine the steady-state limit of ${\ma K}$, the $\vec k$-integrals in (\ref{eq:Kh})-(\ref{eq:Ki}) were evaluated numerically up to $t=10\, 000$, yielding with a four-digit accuracy
\begin{equation}
6\pi \overline{\ma{K}} \simeq
\left( \begin{array}{ccc} 
0.0737  &0 & 0.9436 \\
0 & 0.5766 & 0 \\
0.3425 & 0& 0.3269
\end{array} \right)  \:.
\label{kshear}
\end{equation}
These values are in almost perfect agreement with those obtained by \cite{Miyazaki95a} (their equation (5.27)),  the largest deviation being $0.3\%$. The reason for the tiny differences left between the two sets of coefficients most likely results from truncation errors associated with numerical integration. Note that, despite some similarities, the integrals that appear in the calculation of \cite{Miyazaki95a} and those involved in (\ref{eq:Kh})-(\ref{eq:Ki}) are different, and we do not know how to transform them into each other. Note also that the values of the nonzero components in the first two rows of (\ref{kshear}) differ significantly from those determined by \cite{Harper68}, presumably because of the limited accuracy that they could reach in the numerical integration procedure.\\
In figure \ref{comp_k}, the convergence of the kernel components to their steady-state value is seen to be slow, especially for the ${[\ma{K}]^1}_3$ component. The dimensional time it takes to reach the steady state is of the order of $10\,s^{-1}$ for ${[\ma{K}]^3}_1$, but is typically two orders of magnitude larger for ${[\ma{K}]^1}_3$. For that component, we found that the asymptotic value is approached in a power-law fashion, namely
\begin{equation}
6\pi{[\ma{K}]^1}_3(t) \sim{ 6\pi{[\overline{\ma{K}}]^1}_3} - {C^1}_3 \,t^{-1/2}\:,
\end{equation}
where, according to Maple\textsuperscript{\textregistered}, ${C^1}_3 \approx 1.252$. \\
The slow convergence of the lift components toward their steady-state value has direct consequences on the migration of particles in turbulent flows. In particular, if one plans to examine lateral migration phenomena using a point-particle approach, it is clear from figure \ref{comp_k} that such features may grossly be overestimated if the steady-state values ${[\overline{\ma{K}}]^1}_3$ and ${[\overline{\ma{K}}]^3}_1$ are used instead of the instantaneous values, unless the particle stays in a given vortex (i.e. experiences a given shear rate) during a dimensional time much larger than $s^{-1}$. \\
As pointed out by \cite{Hogg94}, the physical mechanism that produces the lift force corresponding to ${[\overline{\ma{K}}]^3}_1$ may be understood by considering the fluid displaced laterally by the body as it translates along the streamlines of the base flow. In the wake, assuming a positive shear rate, this displaced fluid moves faster (resp. slower) with respect to the body at a given $x_3>0$ (resp. $x_3<0$). At large enough distances from the body, nonlinear advective processes associated with the last two terms in the left-hand side of (\ref{eq_w_adime}) dominate and this asymmetry results in a lateral pressure gradient directed toward negative $x_3$, hence a lift force directed toward positive $x_3$. The mechanism responsible for the lift component associated with ${[\overline{\ma{K}}]^1}_3$ is more subtle because in that case the body translates across the streamlines of the shear flow, and the shear forces the wake to bend. Suppose that the body moves in the direction of increasing velocity, i.e. toward positive $x_3$, and stands at the position where the undisturbed velocity vanishes. Then the fluid contained in the wake experiences a negative transverse velocity that increases with the downstream distance to the body, resulting in a bending of the wake axis toward negative $x_1$. Because of this bending, within a section of the wake perpendicular to its axis, the magnitude of the transverse velocity provided by the undisturbed flow increases with $x_1$. Then, repeating the above argument leads to the conclusion that, at large enough distances from the body, a transverse pressure gradient directed towards $x_1<0$ takes place within each cross section of the wake, resulting in a lateral force on the body directed towards $x_1>0$. As there is no reason for the transverse pressure gradient to be identical in the two situations, it is no surprise that ${[\overline{\ma{K}}]^3}_1\neq{[\overline{\ma{K}}]^1}_3$.

\section{Influence of small inertia effects on the sedimentation of non-spherical particles in a linear flow}
\label{sec:sedimentation}
\noindent It is known that inertia effects make a crucial contribution to the nature of the motion of small, neutrally buoyant non-spherical particles immersed in a shear flow. In particular, \cite{Feng1995} showed numerically that effects of unsteadiness, be they due to the body or the fluid inertia, tend to suppress the periodic oscillations predicted by the quasi-steady approximation. Influence of the body inertia in the case where the particle stands close to a wall also induces dramatic changes because, combined with the wall-particle hydrodynamic interaction, it induces a drift of the particle towards the wall \citep{Gavze1998}. Small-but-finite fluid inertia effects are known to affect the hydrodynamic torque and angular motion in such a way that the marginal stability of the Jeffery orbits of spheroidal particles is broken \citep{subramanian2005,einarsson2015b,candelier2015,Rosen2015,Dabade2016}; these conclusions were recently extended to an arbitrary linear flow field \citep{Marath2018}. Unsteady fluid inertia effects have also been shown to make a significant contribution to the body-shape dependence of the stability exponents of the Jeffery orbits \citep{einarsson2015a}. Most of the above results were obtained by deriving an approximate angular equation of motion for the particle orientation by using a regular first-order perturbation approach with respect to ${\rm Re}_s$. In the case where the particle and fluid densities are different, the particle does not exactly follow the flow, so that a non-negligible slip takes place and the hydrodynamic force is modified by fluid inertia effects at order ${\rm Re}_s^{1/2}$. The question is then that of the influence of the corresponding contributions to the force on the particle path. Addressing this issue requires the translational problem to be solved, which is more challenging than the angular problem, because the corresponding perturbation is singular as discussed in \S \ref{asympt}. In this section, we make use of the developments and results provided earlier in the paper  to consider this question, first for arbitrarily-shaped particles sedimenting in a general linear flow, then in more detail for spheroids immersed in a linear shear flow.
\subsection{General results at ${\mathcal{O}}(\epsilon)$}
\label{general}
In what follows, we implicitly assume that the body density, $\rho_p$, is of the same order as that of the fluid, $\rho_f$, so that the low-Reynolds-number conditions (\ref{Saffman_condition}) may be satisfied within a significant range of fluid viscosities and body sizes. Still assuming $\mbox{St}={\mathcal{O}}(1)$, the body motion is governed by the force and torque balances
\begin{equation}
\frac{\rho_p}{\rho_f} \epsilon^{2} \frac{\mbox{d}}{\mbox{d}t}  \left[\begin{array}{c}
\mathcal{V}_p  \dot{\vec{x}}_p \\
\frac{\mbox{Re}_\omega}{\mbox{Re}_p}\ma{I}_p \cdot \vec{\omega}_p\\
\end{array}\right]  =  \left[
\begin{array}{c}
\vec{f}'^{(0)}\\
\vec{\tau}'^{(0)} \\
\end{array}
\right] + \epsilon 
\left[
\begin{array}{c}
\vec{f}'^{(1)}\\
\vec{\tau}'^{(1)} \\
\end{array}
\right]
 + \left[
\begin{array}{c}
\mathcal{V}_p \left(\frac{\rho_p}{\rho_f}   -1\right) \vec{g}+{\mathcal{O}}(\epsilon^2)\\
{\mathcal{O}}(\epsilon^2) \\
\end{array}
\right]  \:.
\label{eq_mvt_particle}
\end{equation}
Here  $\mathcal{V}_p$ is the non-dimensional volume of the body, $\ma{I}_p$ is its moment-of-inertia tensor, and  $\vec{g}$ denotes the gravitational acceleration normalized by $a(s\nu)^{-1}$. The force and torque $\vec{f}'^{(0)}$ and 
$\vec{\tau}'^{(0)}$ are those corresponding to the Stokes limit   (\ref{eq_wrench}), whereas $\vec{f}'^{(1)}$ and 
$\vec{\tau}'^{(1)}$ are those due to leading-order fluid inertia effects. Terms of ${\mathcal{O}}(\epsilon^2) $ on the right-hand side of (\ref{eq_mvt_particle}) comprise various inertial contributions, including those due to the local acceleration of the undisturbed flow, $\frac{\mbox{D} {\vec{U}}^\infty}{\mbox{D\textit{t}}}=\ma{A}\cdot(\ma{A}\cdot\vec{r})$ (keeping in mind that only steady undisturbed flows are considered in this work). 
 According to (\ref{eq:force2}) and (\ref{eq:torque2}), one has 
\begin{equation}
\left[
\begin{array}{c}
\vec{f}'^{(1)}\\
\vec{\tau}'^{(1)}\\
\end{array}
\right]  = 
-  \left[
\begin{array}{cc}
\ma{M}_{1}(t) & \ma{M}_{2}(t) \\
\ma{M}_{2}^{\sf T}(t) & \ma{M}_{3}(t) \\
\end{array}
\right] \cdot 
\left[
\begin{array}{c}
\int_0^t \ma{K} (t-\tau) \cdot  \tfrac{\rm d}{{\rm \scriptstyle d} \tau}\vec{f}^{(0)}{\rm \scriptstyle d} \tau\\
\vec{0} \\
\end{array}
\right]\,.
\label{fprime}
\end{equation}
Assuming that the slip velocity between the body and fluid is zero at $t=0$, the point force in (\ref{fprime}) has the form
\begin{equation}
\vec{f}^{(0)}(t) = - H(t) \vec{f}'^{(0)}\:,
\label{eq_f_0_H}
\end{equation}
where $H(t)$ denotes the Heaviside function. In keeping with the approximations used throughout the paper, we solve (\ref{eq_mvt_particle}) through an expansion in the small parameter $\epsilon$, seeking the expansion in the form $\dot{\vec{x}}_p=\dot{\vec{x}}_p^{(0)} + \epsilon \dot{\vec{x}}_p^{(1)} + {\mathcal{O}}(\epsilon^2)$, $\vec{\omega}_p=\vec{\omega}_p^{(0)} + \epsilon \vec{\omega}_p^{(1)} +  {\mathcal{O}}(\epsilon^2)$.
To leading order, one has 
\begin{equation}
\left[
\begin{array}{c}
\vec{f}'^{(0)}(\dot{\vec{x}}_p^{(0)},\vec{\omega}_p^{(0)})\:\\
\vec{\tau}'^{(0)} (\dot{\vec{x}}_p^{(0)},\vec{\omega}_p^{(0)}) \\
\end{array}
\right] = 
  \left[
\begin{array}{c}
\mathcal{V}_p  \left(1-\frac{\rho_p}{\rho_f}  \right) \vec{g}\\
\vec{0}\\
\end{array}
\right]  \:,
\label{order0a}
\end{equation}\\
so that
\begin{equation}
\left[\begin{array}{c}
\dot{\vec{x}}_p^{(0)} \\
\vec{\omega}_p^{(0)} \\
\end{array}
\right] = \left[\begin{array}{c}
\vec{U}^\infty \\
\vec{\Omega}^\infty \\
\end{array}
\right]   - \left[
\begin{array}{cc}
\ma{M}_{1}(t) & \ma{M}_{2}(t) \\
\ma{M}_{2}^{\sf T}(t) & \ma{M}_{3}(t) \\
\end{array}
\right]^{-1} \cdot \left[
\begin{array}{c}
\ma{N}_1(t) : \ma{S}^\infty +\mathcal{V}_p  \left(1-\frac{\rho_p}{\rho_f} \right) \vec{g}\\
\ma{N}_2(t) : \ma{S}^\infty\\
\end{array}
\right] \:.
\label{order0}
\end{equation}
Equation (\ref{order0}) describes the gravity-driven settling of the body in the Stokes limit. 
No external force acts at $ {\mathcal{O}}(\epsilon)$, so that 
\begin{equation}
\left[
\begin{array}{c}
\vec{f}'^{(0)}(\dot{\vec{x}}_p^{(1)},\:\vec{\omega}_p^{(1)})\\
\vec{\tau}'^{(0)} (\dot{\vec{x}}_p^{(1)},\:\vec{\omega}_p^{(1)}) \\
\end{array}
\right] = -
\left[
\begin{array}{c}
\vec{f}'^{(1)}(\dot{\vec{x}}_p^{(0)},\:\vec{\omega}_p^{(0)})\\
\vec{\tau}'^{(1)}(\dot{\vec{x}}_p^{(0)},\:\vec{\omega}_p^{(0)}) \\
\end{array}
\right]\,,
\label{order1a}
\end{equation}
which yields
\begin{equation}
\left[\begin{array}{c}
\dot{\vec{x}}_p^{(1)}  \\
\vec{\omega}_p^{(1)} 
\end{array}\right]= 
-\left[\begin{array}{c}
\int_0^t \ma{K}(t-\tau) \cdot 
\frac{\rm \scriptstyle d}{{\rm \scriptstyle d}\tau}  \left(\vec{f}^{(0)}(\dot{\vec{x}}_p^{(0)}(\tau),\vec{\omega}_p^{(0)}(\tau)
\right)
\mbox{d}\tau  \\
\vec{0}
\end{array}\right]\:.
\label{order1}
\end{equation}
Using (\ref{order0a}) and (\ref{eq_f_0_H}) and  noting that $\dot{H}(t) = \delta(t)$, one is finally left with 
\begin{equation} 
\left[
\begin{array}{c}
\dot{\vec{x}}_p^{(1)}  \\
\vec{\omega}_p^{(1)} 
\end{array}
\right]
 = \left[
\begin{array}{c} \mathcal{V}_p\left(1-\frac{\rho_p}{\rho_f}\right) {\ma{{K}}}(t)\cdot  \vec{g}  \\
\vec{0}
\end{array}
\right]
\label{ordre1}
\end{equation}
Hence, only the translational velocity of the body is altered by inertia effects at order $\mbox{Re}_s^{1/2}$, irrespective of the body geometry. In the long-term limit, the kernel $\ma{K}(t)$ tends toward its steady-state value, $\overline{\ma{K}}$. Gathering (\ref{order0}) and (\ref{ordre1}) yields in that limit
\begin{equation}
\begin{split}
\left[\begin{array}{c}
\dot{\vec{x}}_p \\
\vec{\omega}_p \\
\end{array}
\right]   =&  \left[\begin{array}{c}
\vec{U}^\infty \\
\vec{\Omega}^\infty \\
\end{array}
\right]   - \left[
\begin{array}{cc}
\ma{M}_{1}(t) & \ma{M}_{2}(t) \\
\ma{M}_{2}^{\sf T}(t) & \ma{M}_{3}(t) \\
\end{array}
\right]^{-1} \cdot \left[
\begin{array}{c}
\ma{N}_1(t) : \ma{S}^\infty + \mathcal{V}_p  \left(1-\frac{\rho_p}{\rho_f} \right) \vec{g}\\
\ma{N}_2(t) : \ma{S}^\infty\\
\end{array}
\right]  \quad % \Longleftarrow \quad \mbox{Stokes velocity}
\\
& +\epsilon \left[
\begin{array}{c} \mathcal{V}_p\left(1-\frac{\rho_p}{\rho_f}\right) \overline{\ma{K}}\cdot  \vec{g}  \\
\vec{0}
\end{array}
\right] \quad %\Longleftarrow \quad \mbox{Inertia effects}
\end{split}
\label{velocity_long_time}
\end{equation}
Given the structure of the kernel, $\overline{\ma{K}}$ is independent of the shape and initial orientation of the body. Hence, according to (\ref{velocity_long_time}) the same property holds for the long-term translational and angular velocities of the body, or equivalently for the hydrodynamic force acting on it, despite the fact that its rotation may cause unsteadiness at shorter times. This is obviously due to the fact that the external body force and the carrying flow do not depend upon time.  
The remaining question is that of the time it takes for the body to reach such a quasi-steady state.  We shall come back to this in the next subsection.  
\subsection{Sedimentation of prolate and oblate spheroids in a linear shear flow}
\label{sedispher}

 As an application of the above results, we now specialize them to the case of spheroids sedimenting in a linear shear flow. As is well known, a spheroid generally rotates when immersed in a non-uniform flow, owing to the hydrodynamic torque acting on it. However, the resistance tensors $\ma{M}_2$ and $\ma{N}_1$ in (\ref{eq_wrench}) vanish for a spheroidal body, owing to its geometrical symmetries. This implies that there is no coupling between the angular and the translational dynamics of the body, which can therefore be treated separately. Although the body rotation is generally affected by effects of fluid inertia, this alteration only takes place at ${\mathcal{O}}(\epsilon^2)$ \citep{subramanian2005,einarsson2015a,Meibohm2016,candelier2016}. Hence in the ${\mathcal{O}}(\epsilon)$ approximation considered here, the angular velocity of the spheroid is that corresponding to the creeping-flow limit.

In that limit, a small spheroid in a shear flow is known to tumble periodically with an angular velocity obeying \citep{Jeffery22}
\begin{equation}
\vec{\omega}_p  =   \vec{\Omega}^{\infty} + \Lambda  \vec{n} \times \left( \ma{S}^\infty\cdot \vec{n}\right)\,,
\label{eq:jeffery}
\end{equation}
where $\Lambda=(\lambda^2-1)(\lambda^2+1)^{-1}$ is a shape parameter that depends on the body aspect ratio, $\lambda$, which is the ratio of the body length along the symmetry axis to that of its equatorial diameter. The normalizing length, $a$, considered so far is taken to be the half-length of the major semi-axis, while $b$ is the half-length of the minor semi axis, so that $\lambda=a/b$ (resp. $b/a$) for a prolate (resp. an oblate) spheroid.
 The kinematic equation $\tfrac{{\rm d}}{{\rm d}t}{\vec n} = \vec \omega_p \times\vec n $ governing the evolution of the orientation of the body symmetry axis (see figure \ref{Diph_fig1}), has an infinite number of marginally stable periodic solutions, commonly known as Jeffery orbits.  Here we assume that the symmetry vector $\vec n$ tumbles within the $(\vec e_1,\vec e_3)$ plane where the shear flow takes place (this is the orbit expected to produce the largest unsteadiness). With this choice, the angular velocity of the spheroid is related to the angle $\theta(t)$ made by the spheroid axis with the streamlines of the shear flow through $\vec{\omega}_p = \dot{\theta}(t) \vec{e}_2$, and $ \theta(t)$  obeys
\begin{equation}
\dot{\theta}(t) = \frac{1}{2}+\Lambda \Big[ \frac{1}{2} -\cos(\theta(t))^2\Big] \:.
\end{equation}
This ordinary differential equation has a periodic solution \citep{Jeffery22}, characterized by a period $T_J=(2\pi)/\sqrt{1-\Lambda^2}$.  As a result, $\vec{n}$ rotates within the ($\vec{e}_1$, $\vec{e}_3$) plane according to   
\begin{equation}
\vec{n}(t) =  \vec{e}_1\cos\theta(t) +  \vec{e}_3 \sin\theta(t)\:. 
\label{eq:n}
\end{equation}
This periodic angular motion acts as an unsteady disturbance for the translational problem. We assume $\theta(0)=0$, i.e. the symmetry axis of the spheroid is initially aligned with the streamlines of the base flow.  Following (\ref{eq_wrench}), one has
\begin{equation}
\vec{f}^{(0)} = \ma{M}_1(t) \cdot \vec{u}_s(t) \,,
\end{equation}
where $\vec{u}_s(t)=\vec U^\infty(\vec{x}_p(t))-\vec{\dot x}_p(t)$ denotes the instantaneous slip between the body and fluid. The resistance tensor is known to be diagonal in the principal axes of the spheroid \citep{kim1991}, so that 
\begin{equation}
\ma{M}_1= M^{\parallel} \vec{n}\vec{n}+ M^{\bot} \left( \ma{1} - \vec{n}\vec{n}\right)\:,
\label{eq:tensor_M}
\end{equation}
where 
the components  $M^{\parallel}$ and $M^{\bot}$ depend on the aspect ratio, $\lambda$. 
 For a prolate spheroid ($\lambda>1$) they are
 \begin{equation}
M^{\parallel} \!=\! \tfrac{8}{3 \lambda} \frac{6\pi }{{ \frac{ -2 \lambda}{\lambda^2-1} +\tfrac{2 \lambda^2-1}{(\lambda^2-1)^{3/2}} \ln \tfrac{\lambda+ \sqrt{\lambda^2-1}}{\lambda-\sqrt{\lambda^2-1}}}}
\,,
M^{\bot} \!=\! \tfrac{8}{3 \lambda} \frac{6\pi }{{ \frac{ \lambda}{\lambda^2-1} +\frac{2 \lambda^2-3}{(\lambda^2-1)^{3/2}} \ln \lambda+ \sqrt{\lambda^2-1}}}\:,
\end{equation}
whereas for an oblate spheroid ($\lambda<1$) one has
\begin{equation}
M^{\parallel} \!=\! \tfrac{8}{3 } 
\frac{6\pi }{{ \frac{ 2 \lambda}{1-\lambda^2}+\frac{2 (1-2 \lambda^2)}{(1-\lambda^2)^{3/2}}
\tan^{-1}\frac{\sqrt{1-\lambda^2}}{\lambda}}}\,,
M^{\bot} \!=\! \tfrac{8}{3 } \frac{6\pi }{{ -\frac{\lambda}{1-\lambda^2}-\frac{(2 \lambda^2-3)}{(1-\lambda^2)^{3/2}}
\sin^{-1}\sqrt{1-\lambda^2}}}\:.
\end{equation}
Due to the body rotation, the orientation vector $\vec{n}(t)$ depends upon time and so do the components of $\ma{M}_1$ in the laboratory frame.\\

To reveal the influence of small inertia effects on the body dynamics, we numerically integrated (\ref{eq_mvt_particle})-(\ref{fprime}) with the following dimensional parameters: $a=1\: \mbox{mm}$, $\nu= 10^{-4}\: \mbox{m}^2/\mbox{s}$, $s=10 \,\mbox{s}^{-1}$, and ${\rho_p}/{\rho_f} =1.5$. These  parameters imply $\epsilon=0.316$, so that predictions provided by the asymptotic approach are expected to be at least qualitatively valid. We consider four distinct spheroid aspect ratios, namely $\lambda=1/10,\,1/2,\, 2$ and $10$.  Gravity is set in the form $\vec{g}=-g\vec{e}_3$, so that spheroids are settling along the shear. Integration of (\ref{eq_mvt_particle})-(\ref{fprime}) is achieved using a method inspired from \cite{daitche2013}, with the history integral in (\ref{fprime}) evaluated using an implicit scheme.
\begin{figure}
\begin{center}
\hspace{-3mm}\includegraphics[width=200 pt]{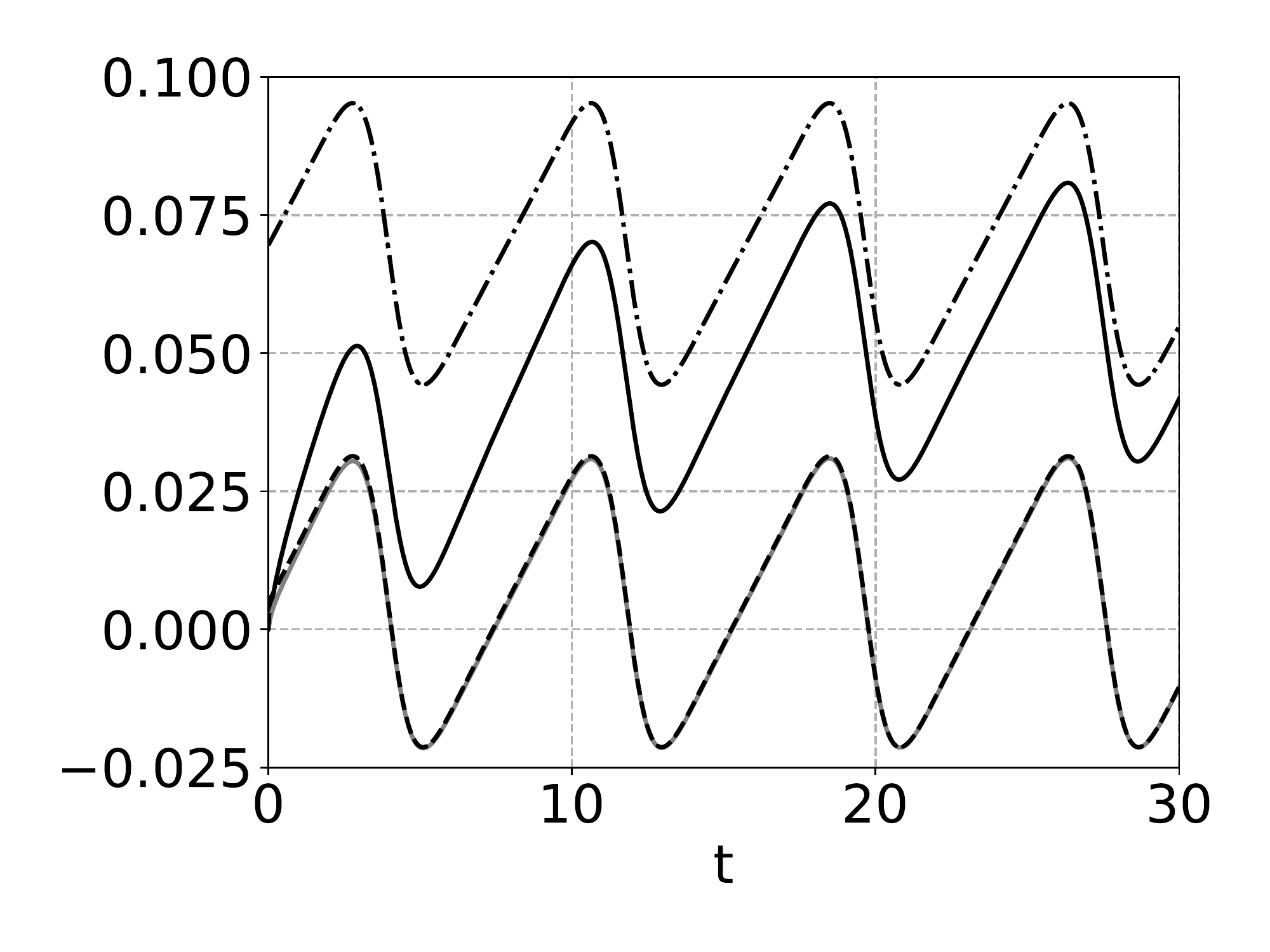}
\hspace{-3mm}\includegraphics[width=200 pt]{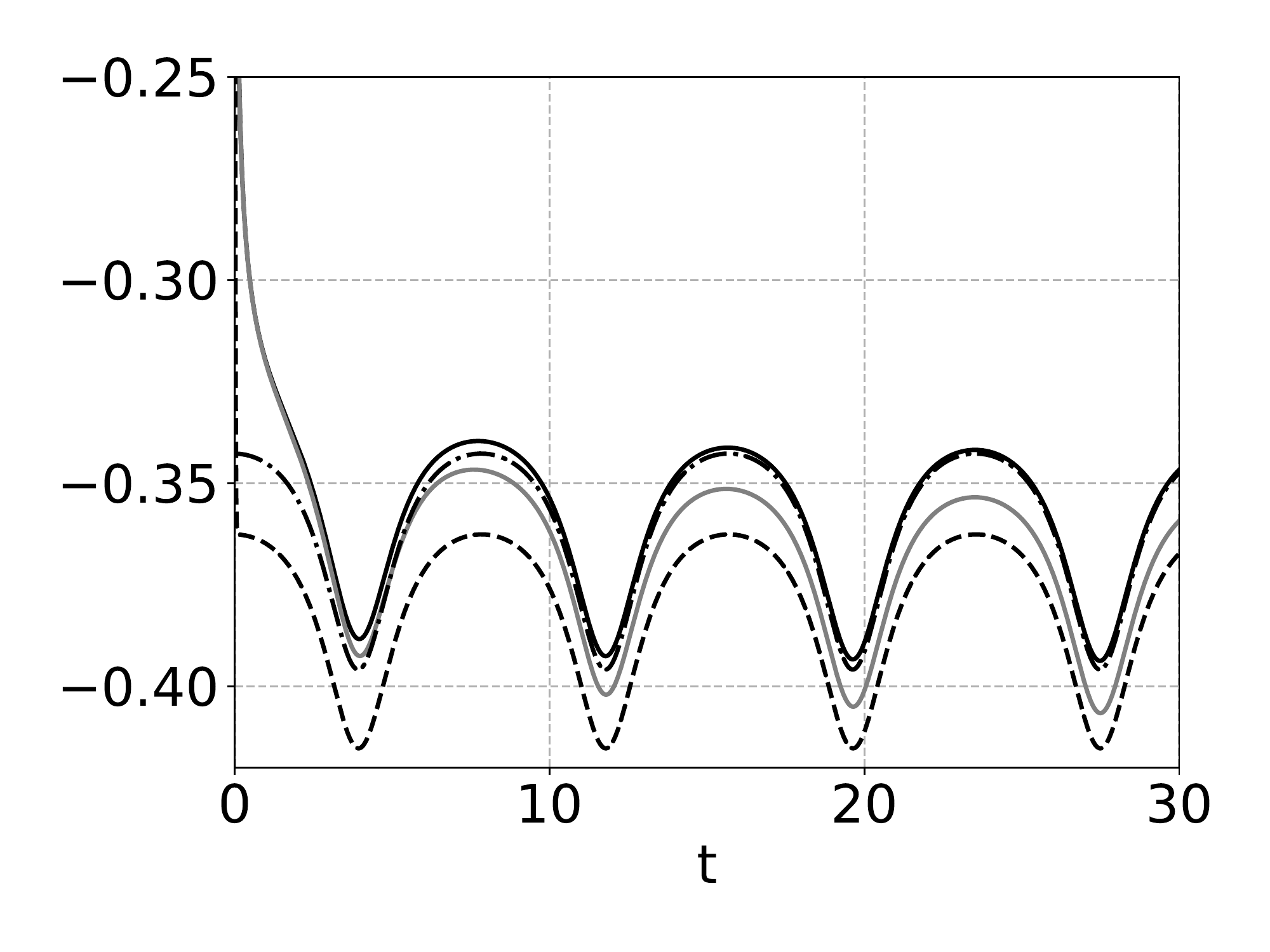}
\begin{flushleft}
\vspace{-16.5mm}
\hspace{50mm}\large{$(a)$}\hspace{61.5mm}\large{$(b)$}
\end{flushleft}
\vspace{9mm}
\end{center}
\caption{Evolution of $(a)$ the $\vec{e}_1$-component, and $(b)$ the $\vec{e}_3$-component of the slip velocity, $\vec{u}_s$, of a prolate spheroid with aspect ratio $\lambda=2$, as predicted using different approximations. Black line: present unsteady theory; dash-dotted line: present quasi-steady theory; dashed line: prediction based on the Stokes quasi-steady drag; grey line:  prediction based on the sum of the Stokes quasi-steady drag and the Basset-Boussinesq force.}
\label{fig_us0_prolate_2}
\end{figure}
\begin{figure}
\begin{center}
\hspace{-3mm}\includegraphics[width=200 pt]{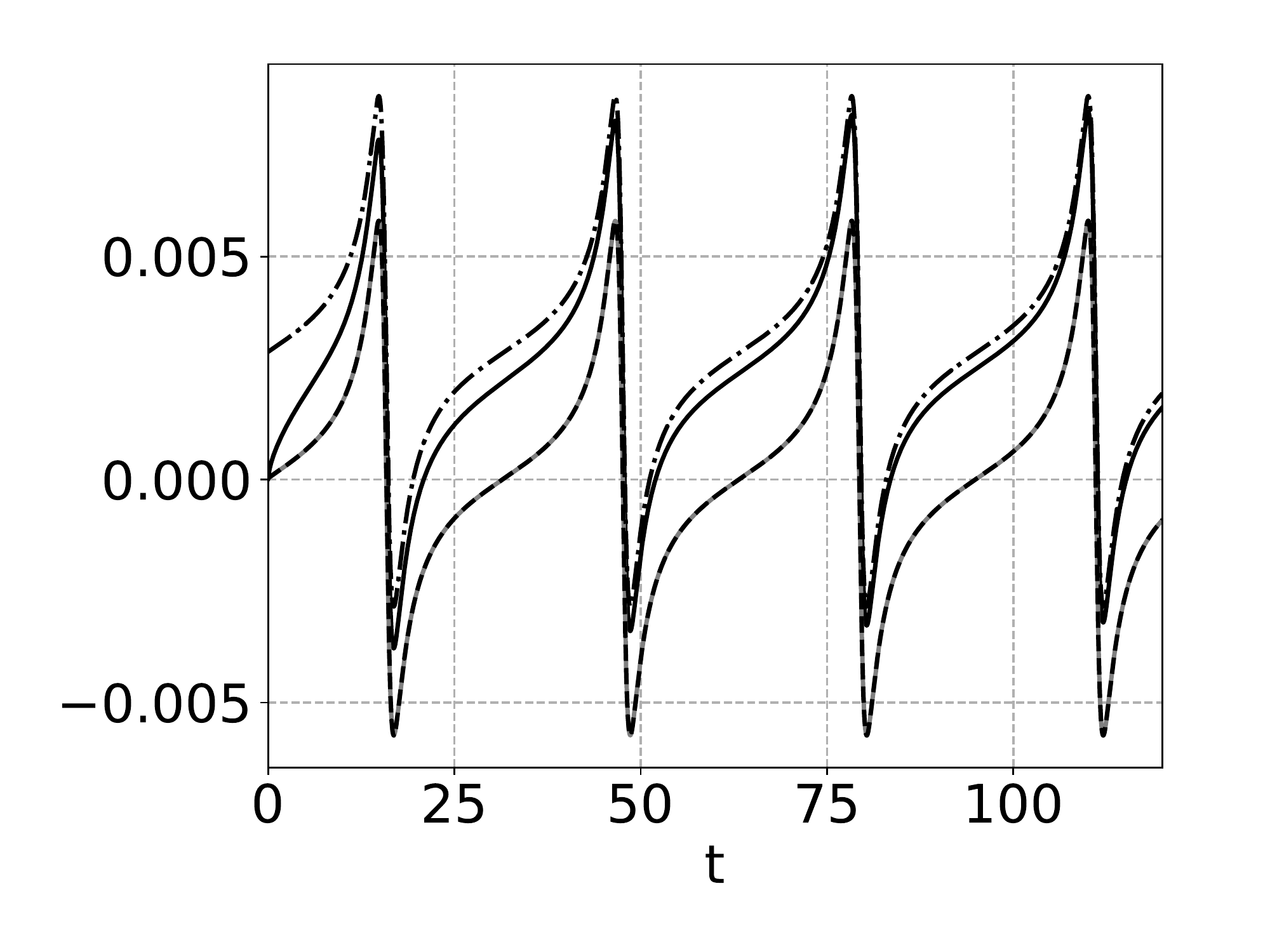}\hspace{-3mm}\includegraphics[width=200 pt]{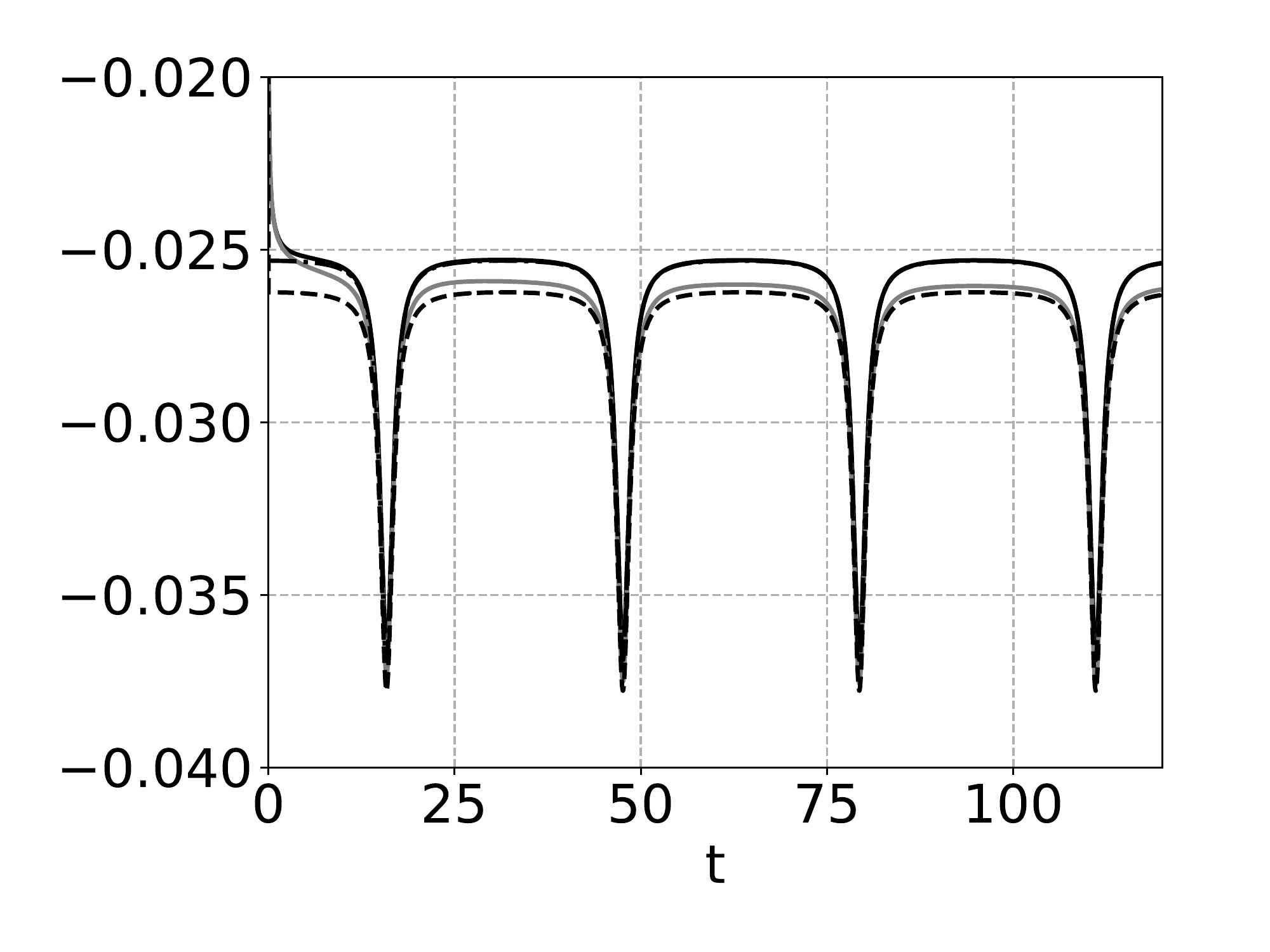}
\begin{flushleft}
\vspace{-16.5mm}
\hspace{48mm}\large{$(a)$}\hspace{64mm}\large{$(b)$}
\end{flushleft}
\vspace{9mm}
\end{center}
\caption{Same as figure \ref{fig_us0_prolate_2} for a prolate spheroid with aspect ratio $\lambda=10$.}
\label{fig_us0_prolate_10}
\end{figure}
The results of this integration are displayed in figures \ref{fig_us0_prolate_2} to \ref{fig_us0_oblate_01}. Let us first comment on the predictions in which inertial corrections are ignored. Clearly, predictions of the horizontal slip component, $\vec{u}_s\cdot\vec{e}_1$, obtained by considering only the Stokes quasi-steady drag (dashed lines) and those in which the Basset-Boussinesq `history' force is also taken into account (grey lines) are indiscernible, indicating that the relative acceleration between the body and the fluid does not play any role on that component. In contrast the two predictions for the vertical slip component, $\vec{u}_s\cdot\vec{e}_3$, differ significantly all along the body path, especially when the body aspect ratio is of order unity (figures \ref{fig_us0_prolate_2} and \ref{fig_us0_oblate_05}). It is worth noting that the time-averaged horizontal slip is not strictly zero, as may especially be inferred from the above two figures. Although surprising at first glance, the small positive value of the time-averaged horizontal slip velocity may be understood by considering the zeroth-order horizontal force balance in (\ref{eq_mvt_particle}). Expressed with respect to $(\dot{\vec{x}}_p-\vec{U}^\infty)\cdot\vec{e}_1$, this force balance provides a first-order differential equation governing the horizontal slip component. It involves the source term $-\frac{\rho_p}{\rho_f} \epsilon^{2}\dot{\vec{U}}^\infty$, where $\dot{\vec{U}}^\infty=(\dot{\vec{x}}_p\cdot\vec{e}_3)\vec{e}_1$ denotes the time rate-of-change of the fluid velocity along the body path. Hence, as the body moves in the vertical direction, it `sees' a time-varying horizontal background velocity, which makes the average horizontal slip non strictly zero.\\
Let us now turn to the influence of inertial corrections. In all four cases, it is seen that these corrections result in a large positive shift of the horizontal slip. In particular, this slip component is now positive for all times, i.e. the spheroid drifts in the $\vec{e}_1$-direction, for a spheroid with $\lambda=2$ (figure \ref{fig_us0_prolate_2}). The same feature is observed with the oblate spheroid corresponding to $\lambda=0.5$  (figure \ref{fig_us0_oblate_05}), except at very short imes. There is also a slight reduction of the vertical slip velocity, which corresponds to the increase of the drag force associated with the positive diagonal component ${[\overline{\ma{K}}]^3}_3$ in (\ref{shortt}) and (\ref{kshear}). This slip component is found to converge within a  few units of dimensionless time. The convergence of the horizontal slip component is much slower. This is no surprise since this component directly depends on $[\ma{K}]^1_3(t)$, which was found to converge very slowly toward its steady-state value in figure \ref{comp_k}. Moreover, comparing panels $(a)$ in figures \ref{fig_us0_prolate_2} and \ref{fig_us0_prolate_10} on the one hand, and in figures \ref{fig_us0_oblate_05} and \ref{fig_us0_oblate_01} on the other hand reveals that effects of unsteadiness in the kernel $\ma{K}$ manifest themselves over much longer times for spheroids with ${\mathcal{O}}(1)$ aspect ratios. Considering the steady-state expression of the kernel overestimates the horizontal slip by as much as $50\%$ for both $\lambda=1/2$ and $\lambda=2$, even after several tens of time units.\vspace{2mm}\\
The above examples shed light on the importance of inertial corrections to the hydrodynamic force on the path of spheroids sedimenting in a shear flow. In particular, they show that the horizontal component of the slip velocity cannot realistically be predicted on the basis of the forces derived in the creeping-flow approximation, be it quasi-steady or fully unsteady. Moreover they demonstrate that, owing to the slow convergence of the inertial kernels, large errors can be made in the prediction of this slip component if the quasi-steady approximation of the inertial corrections to the force is used in place of their time-dependent expression.
\begin{figure}
\begin{center}
\hspace{-3mm}\includegraphics[width=200 pt]{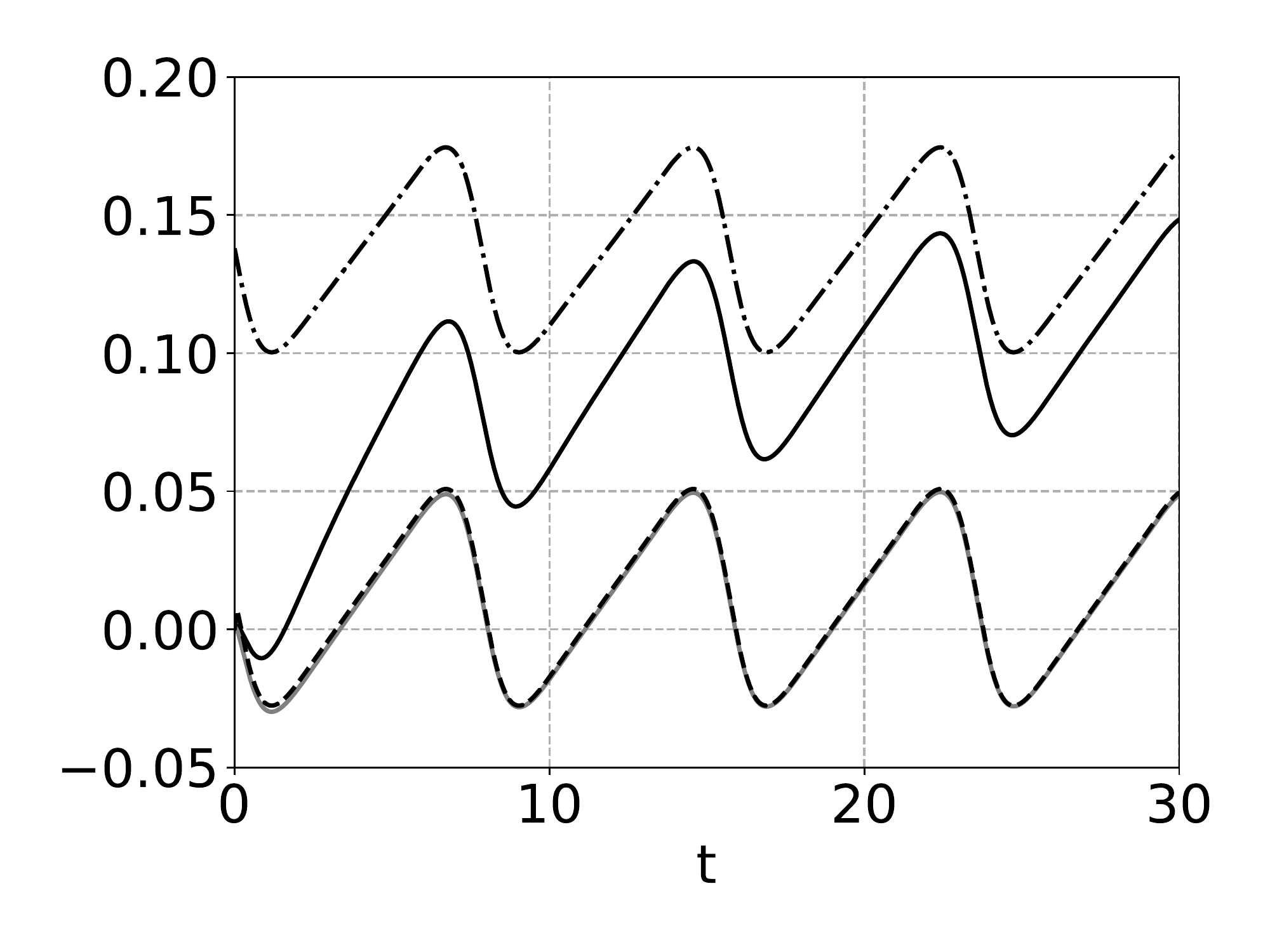}\hspace{-3mm}\includegraphics[width=200 pt]{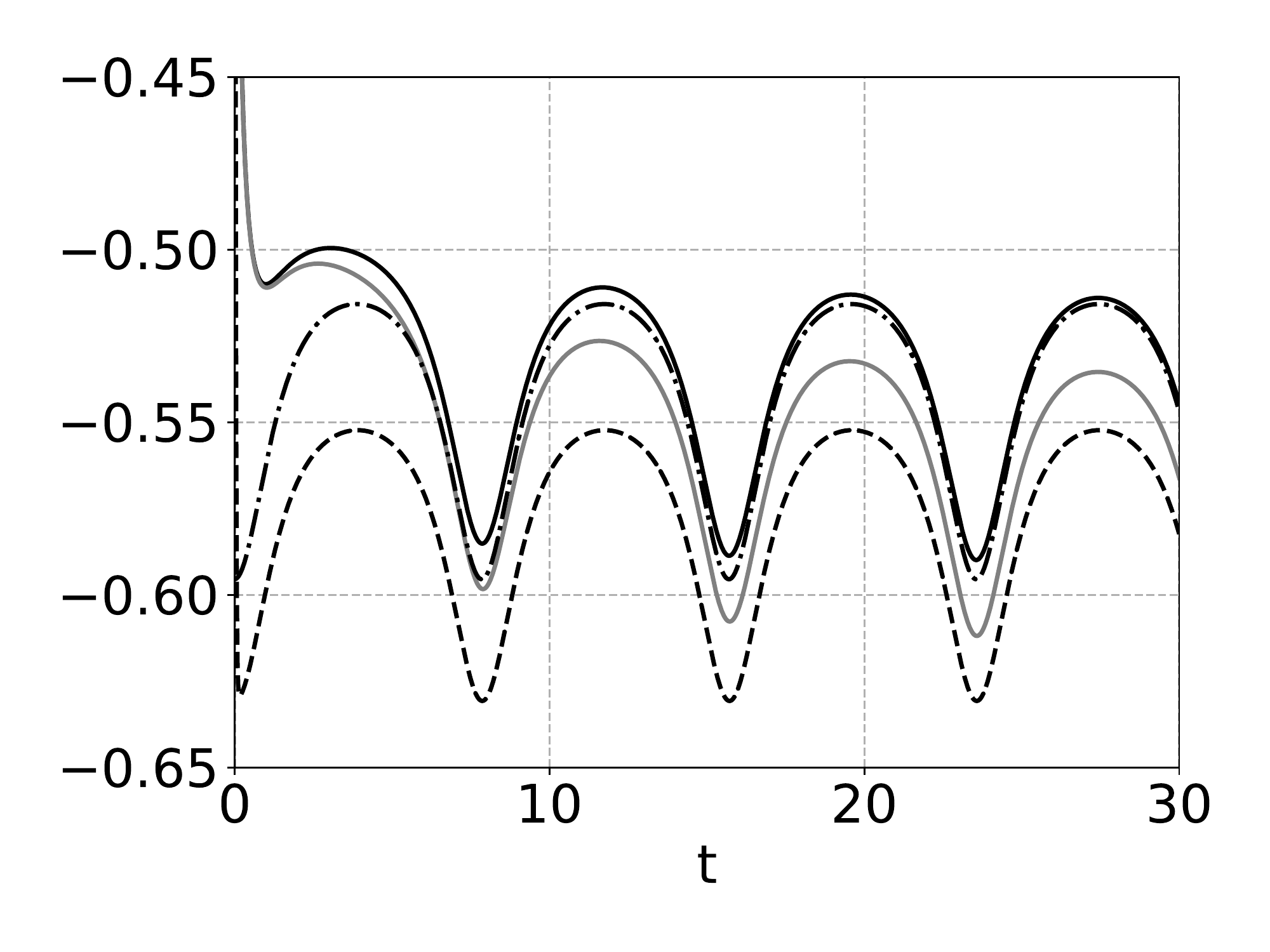}
\begin{flushleft}
\vspace{-16.5mm}
\hspace{55mm}\large{$(a)$}\hspace{61.5mm}\large{$(b)$}
\end{flushleft}
\vspace{9mm}
\end{center}
\caption{Same as figure \ref{fig_us0_prolate_2} for an oblate spheroid with aspect ratio $\lambda=1/2$.}
\label{fig_us0_oblate_05}
\end{figure}
\begin{figure}
\begin{center}
\hspace{-3mm}\includegraphics[width=200 pt]{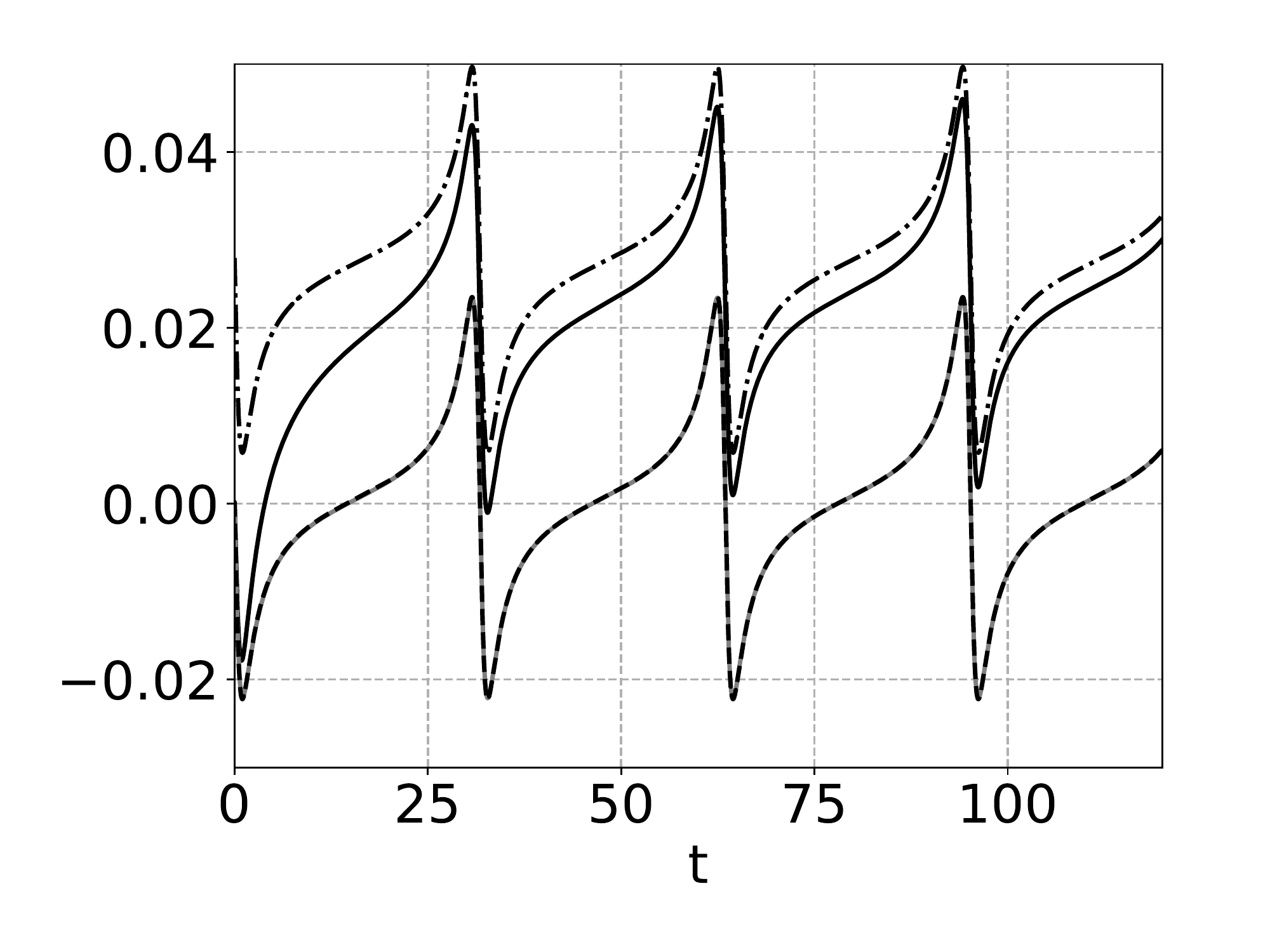}\hspace{-3mm}\includegraphics[width=200 pt]{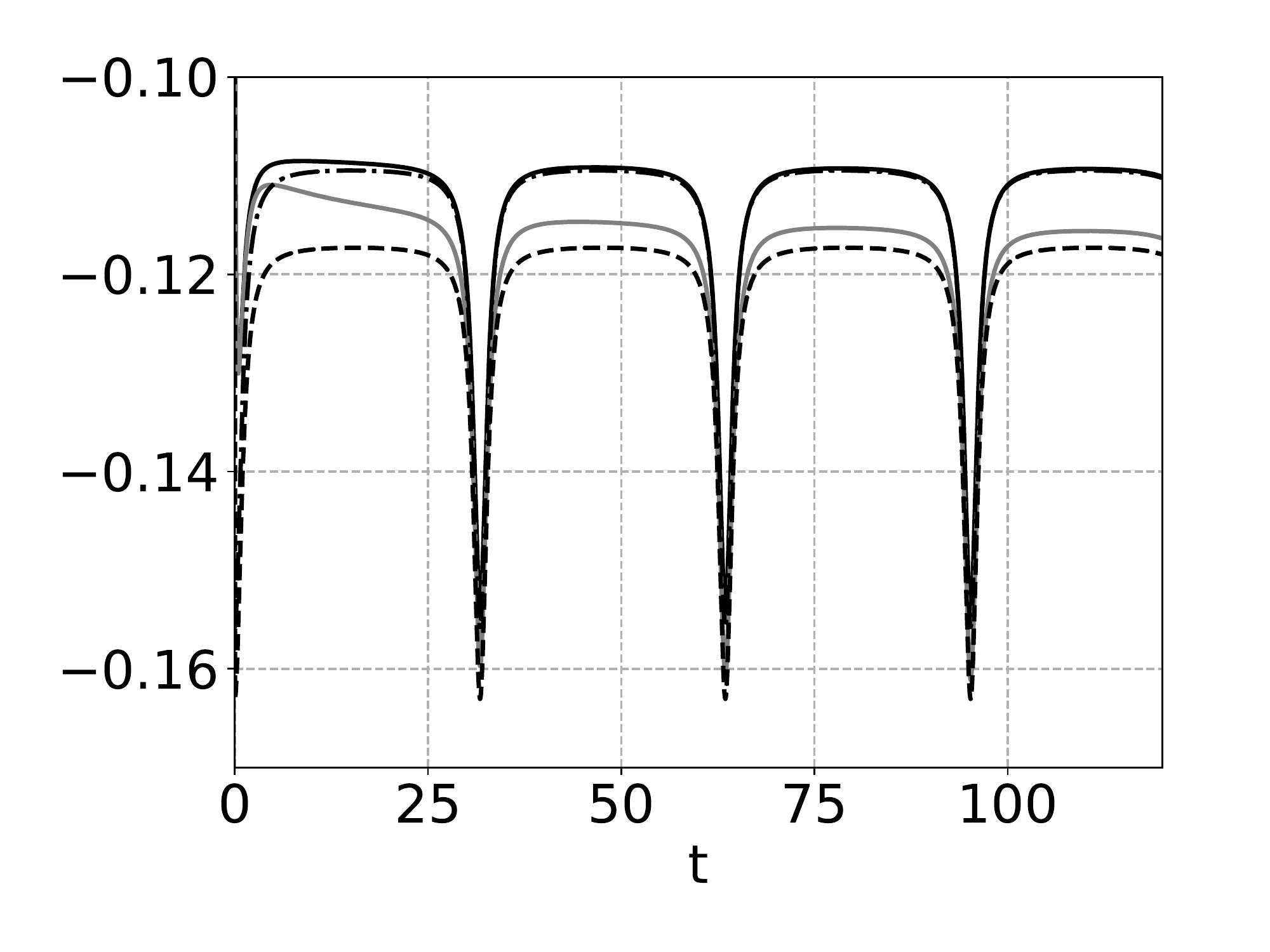}
\begin{flushleft}
\vspace{-16.5mm}
\hspace{53mm}\large{$(a)$}\hspace{64mm}\large{$(b)$}
\end{flushleft}
\vspace{9mm}
\end{center}
\caption{Same as figure \ref{fig_us0_prolate_2} for an oblate spheroid with aspect ratio $\lambda=1/10$.}
\label{fig_us0_oblate_01}
\end{figure}

\section{Summary and prospects}
\label{sec:conc}
\noindent In this paper, motivated by applications to turbulent particle-laden flows, we developed a generic methodology aimed at determining the leading-order inertial corrections to the instantaneous force and torque acting on an arbitrarily-shaped rigid body moving with a time-dependent slip velocity in a quasi-steady linear flow field. We carried out the corresponding developments in the framework of the MAE approach, under the assumption that the variations of the undisturbed velocity at the body scale are much larger than the slip velocity, so that inertial effects due to the latter may be neglected at leading order.  The key of the success was to express the flow disturbance in a non-orthogonal co-moving coordinate system that reduces the initial set of partial differential equations governing the disturbance problem in Fourier space to a set of ordinary differential equations that are much more easily solved whatever the nature of the background linear flow. The solution of this differential problem was obtained in the form of a closed convolution kernel, thanks to the use of Magnus expansions. The above idea is in essence similar to that used by \cite{Miyazaki95a} in the framework of the IF formulation. However it is somewhat hidden in their work, where it appears only through the use of time-dependent wavevectors during the step when the solution of the disturbance problem is sought in Fourier space. Because of these differences, the kernels provided by the two approaches exhibit a quite different mathematical structure, although they must yield identical predictions once integrated for any specific time variation of the slip velocity. \\
We proved the versatility of our approach by computing explicitly the kernel in the case of a body moving in a planar flow corresponding to a solid body rotation, a planar elongation or a uniform shear. In the first case, as expected, we recovered the kernel derived by \cite{Candelier08} using a change of reference frame. In the uniform shear configuration, all kernel components agree very well with those computed by \cite{Miyazaki95a} in both the short- and long-time limits, thus providing a stringent validation of the various steps involved in the present procedure. We actually computed the kernel for arbitrary times, which revealed in particular that some components require a much longer time than others to reach their steady-state value. Few results are available for the purely elongational case which we also considered. Our results recover the short-time behaviour predicted by \cite{Bedeaux87}. At longer time, we noticed an unexpected behaviour. While the kernel component corresponding to the extensional direction converges gradually towards its steady-state value (which differs from that predicted by \cite{Drew78}), the component corresponding to the compressional direction changes sign, implying that long-term inertial effects tend to decrease the drag force in that direction. Unfortunately, we have not yet computed the steady-state value of this component, due to technical difficulties encountered in the numerical integrations. We are still working on this issue.  \\
As shown in \S\ref{general}, once the kernel is determined, the MAE approach allows the leading-order inertial force and torque corrections on a non-spherical body to be evaluated in a straightforward manner, provided the body's resistance tensors are known (the same remark applies to drops and bubbles with a prescribed shape, for which the appropriate kernel may be directly deduced from that of the corresponding solid body by applying the argument developed by \cite{Legendre97}). We illustrated this in \S\ref{sedispher} by considering the sedimentation of spheroids, the rotational dynamics of which is unaffected by inertial effects at the order considered here. In contrast, we found that the horizontal component of their slip velocity is dramatically enhanced by these effects. Consequently, neither the Stokes approximation nor the refined approximation including the Basset-Boussinesq history force provides a reasonable prediction of this characteristics. Although closer to the actual evolution, the prediction based on the quasi-steady limit of the inertial corrections is also poorly accurate, especially for spheroids with moderate aspect ratios, owing to the aforementioned slow convergence of some of the kernel components. \vspace{2mm}\\
The results provided in this paper represent a first step toward a rational extension of the Basset-Boussinesq-Oseen approximation describing the unsteady motion of a small rigid particle to situations involving small-but-nonzero inertial effects due to the carrying flow. Such an extension is key to improving the determination of the forces and torques that govern the motion of particles immersed in a turbulent flow. To progress toward this objective, we now plan to extend present results in several directions. Our first objective remains to obtain the expression of the time-dependent kernel in a general steady linear flow characterized by an arbitrary traceless velocity gradient tensor, $\ma{A}$. To maximize the usefulness of the outcome in terms of applications, we shall seek the components of that kernel in a general form involving explicitly the ${\ma{A}^i}_j$ components, in such a way that the inertial corrections can be straightforwardly computed in any linear flow once $\ma{A}$ is known. By proceeding in this manner, all results corresponding to the canonical flows considered here will be recovered as special cases. Analytical forms of the kernel will certainly be limited to the short- and long-times limits and we will have to develop approximate fits to provide expressions valid for arbitrary times. \vspace{2mm}\\
Then, two central assumptions extensively used in the present work will have to be removed. First of all, we will have to consider that the deformation tensor, $\ma{A}$, may be time-dependent. This is essential for the prediction of particle motion in turbulent flow, owing to the aforementioned slow convergence of several kernel components. Indeed, the turnover time of small-scale eddies, which have the largest velocity gradients, is too short to allow these kernel components to reach their steady state value within a time interval during which the carrying flow may be considered frozen. For instance, only eddies larger than the Taylor micro-scale have a turnover time larger than the viscous time $\eta_k^2/\nu$ corresponding to particles with a characteristic size of the order of the Kolmogorov length scale, $\eta_k$. Additional technical difficulties are expected with time-dependent flows because (\ref{eq:Exp_A}) is not valid any more and a new term appears in (\ref{eq_NS4}). Moreover, when $\ma{A}$ depends upon time, the solution of (\ref{eq:fundamental_pb}) does not depend on the time lag $t-\tau$ only, so that the final expression of the inertial corrections does not simplify to a convolution product any more.\\
Second, in the same spirit as the extension carried out by \cite{McLaughlin91} in a pure shear flow for the quasi-steady Saffman lift force, we will have to go beyond Saffman's condition (\ref{Saffman_condition}) in a general linear flow. For this, we must allow the two Reynolds numbers, $\mbox{Re}_s$ and $\mbox{Re}_p$, to be of similar magnitude and examine how the kernel varies with the ratio $\epsilon_{sp}=\sqrt{\mbox{Re}_s}/\mbox{Re}_p$ comparing the Oseen length scale, $\ell_o=\mbox{Re}_p^{-1}$, to the Saffman length scale, $\ell_s$. In this way we shall cover time-dependent situations in which dominant advective corrections are due to shearing effects (as in the present paper) as well as situations in which Oseen-like effects dominate. The limit $\epsilon_{sp}\rightarrow0$ corresponds to the time-dependent problem considered by \cite{Lovalenti1993a}, who showed that advective effects drastically reduce the long-term magnitude of the `history' force because a vorticity disturbance resulting from a change in the slip velocity is more efficiently removed from the body's vicinity by these effects than by viscous diffusion once it has entered the Oseen region of the body-induced flow. For instance `history' effects decay as $t^{-2}$ at large times in the case of a sudden start of the body \citep{Sano81,Lovalenti1993a}, in contrast to the $t^{-1/2}$ behaviour predicted by the Basset-Boussinesq kernel. Technically, considering finite values of $\epsilon_{sp}$ amounts to replacing the term $\ma{A} \cdot \vec{\hat{w}} $ in (\ref{eq_w_hat}) by $\{\ma{A}-\epsilon_{sp}^{-1}\mbox{Re}_s^{-1/2}(\mathtt{i}\vec{k} \cdot \vec{u}_s)\ma{1}\}\cdot  \vec{\hat{w}} $. Since $\vec{u}_s$ is generally time-dependent, the extra difficulty is similar to that encountered with time-dependent velocity gradients.\\
The above extensions will involve a substantial amount of numerics, since solving the disturbance equation in Fourier space by hand or with the help of a symbolic computation software is only possible in very specific cases. More precisely, the solution may be obtained in this way at large $\vec{k}$, but in most cases the ordinary differential equation needs to be solved numerically in the small-$\vec{k}$ range which is the one that provides the leading contributions to the long-term kernel.\vspace{2mm}\\
As a last extension, we wish to determine the second-order inertial corrections, at least in selected situations (e.g. spheroidal bodies, canonical flows), for several reasons. First, both the translational and the rotational dynamics are modified by inertial effects at ${\mathcal{O}}(\epsilon^2)$, making this order of approximation relevant to obtain a nearly complete view of the influence of small-but-finite inertial effects on the dynamics of particles in turbulent flows, as well as on the rheology of sheared suspensions. It is also an order of approximation where couplings between translation and rotation may happen, even for symmetric body shapes for which the coupling resistance tensor $\ma{M}_2$ is zero. This is for instance the case of the ${\mathcal{O}}(\mbox{Re}_\omega)$ lift force experienced by a spinning sphere translating in a fluid at rest \citep{Rubinow1961}, and an ${\mathcal{O}}(\epsilon^2)$ expansion should capture this effect. Last, added-mass effects resulting from the differential acceleration between the body and the carrying flow are also of ${\mathcal{O}}(\epsilon^2)$. This is why these effects were not captured in the kernels computed in sections \ref{bidir} and \ref{sec:kernel}, unlike the ${\mathcal{O}}(\epsilon)$ `history' effects. Expanding the solution of the disturbance problem up to ${\mathcal{O}}(\epsilon^2)$-terms in linear flows would allow us to clarify the expression of the differential acceleration involved in the added-mass force: although it is known that this contribution is proportional to the difference between the Lagrangian acceleration $\frac{\mbox{D} {\vec{U}}^\infty}{\mbox{D\textit{t}}}$ and the body acceleration $\frac{\mbox{d}\dot{\vec{x}}_p}{\mbox{dt}}$ in an inviscid flow \citep{Taylor1928, Auton1988}, the counterpart in the regime of low-but-finite Reynolds numbers is unknown. Clarifying this issue and gathering all inertial effects in a rational way up to ${\mathcal{O}}(\epsilon^2)$ would represent a major extension of the Basset-Boussinesq-Oseen equation, even in the simplest case of a sphere, since only \textit{ad hoc} extensions of this equation towards the inertial regime are available so far in nonuniform flows.\\

{\em  BM was supported by  Vetenskapsr\aa{}det [grant 2017-03865], Formas [grant number 2014-585],  and by the grant \lq Bottlenecks for particle growth in turbulent aerosols\rq{} from the Knut and Alice Wallenberg Foundation, Dnr. KAW 2014.0048.}

%%%%%%%%%%%%%%%%%%%%%%%%%%%%%%%%%%%%%%%%%%%%%%%%%%%%%%%%%%%%%%%%%%%%%%%%%%%%%%%%%%%%%%%%
%                                                           APPENDIX                                                                                                            %
%%%%%%%%%%%%%%%%%%%%%%%%%%%%%%%%%%%%%%%%%%%%%%%%%%%%%%%%%%%%%%%%%%%%%%%%%%%%%%%%%%%%%%%%%

\clearpage
\appendix

 \section{Solving the disturbance equation in the co-moving reference frame}
 \label{AppendixA}

\subsection{Change of coordinates}

Equation (\ref{eq:derivative_transform}) links the partial derivative of the velocity with respect to time evaluated at fixed $r^i$ with that evaluated at fixed $R^i$. To derive it, we first compute the (total) time rate-of-change of the velocity  while following the motion of a $R^i$-coordinate position.  Along the corresponding path, the velocity $\vec{w}$ may be written either in the Cartesian basis, $\vec{e}_i$, 
\begin{equation}
\vec{w} = w^i \Big(\vec{r}(R^i,\: t), \:t \Big) \vec{e}_i\,,
\end{equation}
or in the co-moving basis, $\vec{E}_i$,
\begin{equation}
 \vec{w} =  W^i (R^i,\:t) \vec{E}_i (t)\:.
\end{equation}
The time rate-of-change of $\vec{w}$ is then either
\begin{equation}
\frac{\mbox{d} \vec{w}}{\mbox{d}t} =  \frac{\partial (w^i(r^i(t), t) \vec{e}_i )}{\partial t}\big|_{r^i} + \vec{v}\cdot \boldsymbol{\nabla}(w^i(r^i(t), t) \vec{e}_i) \:
\label{e1}
\end{equation}
in an Eulerian-like description (with $\vec{v}=( \partial  \vec{r}/\partial t) \big|_{R^j}$), or 
\begin{equation}
\frac{\mbox{d} \vec{w}}{\mbox{d}t} = \frac{\partial (W^i \vec{E}_i)}{\partial t}\big|_{R^i} 
\label{e2}
\end{equation}
in a Lagrangian-like approach. 
 From (\ref{e1}) and (\ref{e2}), it is immediate to conclude that
\begin{equation}
\frac{\partial (w^i(r^i(t), t) \vec{e}_i )}{\partial t}\big|_{r^i} + \vec{v} \cdot \boldsymbol{\nabla}(w^i(r^i(t), t) \vec{e}_i) = \frac{\partial (W^i \vec{E}_i)}{\partial t}\big|_{R^i} \,,
\end{equation} 
which proves (\ref{eq:derivative_transform}).

\subsection{Solution in  the co-moving basis }

Index notation was used throughout section \ref{sec:coords} to avoid ambiguity in the derivation of (\ref{eq_NS4}). However, in order to solve effectively this equation, it is appropriate to switch to matrix notation.  To do so, contravariant components of vectors, such as $W^i$, are stored in column vectors, while covariant components, such as $K_i$, are stored in line vectors. Components of the metric tensor, or of tensors $\ma{F}$ and $\ma{A}$, are stored in matrix form.  It must be pointed out that the mathematical objects we are dealing with in what follows are not necessarily tensors but may be simple matrices. 

Using such conventions and after the pressure has been eliminated with the help of the divergence-free condition, the Fourier transform of (\ref{eq_NS4}) may be cast in the form
\begin{equation}
\frac{\partial \hat{\vec{W}}}{\partial t}\big|_{K_i} = \ma{H} \cdot \hat{\vec{W}} - K^2 \hat{\vec{W}} + K^2 \hat{\vec{T}}^{(0)}\,,
\label{eq_NS6}
\end{equation}
with
\begin{equation}
K^2(\vec{K},t) = \vec{K} \cdot \ma{R}(t) \cdot \vec{K}^{\sf T} \:,
\label{eq:K2}
\end{equation}
\begin{equation}
\hat{\vec{T}}^{(0)}(\vec{K},t) = \frac{1}{K^2} \left( \ma{1}  + \frac{(\ma{R}(t) \cdot \vec{K}^{\sf T} ) \vec{K}}{K^2}\right)\cdot \vec{F}^{(0)}(t)\,,
\label{eq:Stokeslet_curvilinear}
\end{equation}
\\
and
\begin{equation}
\ma{H}(\vec{K},t) = -2 \ma{A}  + 2  \frac{( \ma{R}(t) \cdot \vec{K}^{\sf T}) (\vec{K} \cdot \ma{A})}{K^2}\:,
\end{equation}
\vspace{2mm}\\
where, as defined in section \ref{sec:coords}, $\ma{R}$ is the inverse of the metric tensor $\ma{g}$ associated with the coordinate transformation. In order to solve (\ref{eq_NS6}), we first need to determine the solution 
of the homogeneous problem, which we denote by ${\hat{\vec{W}}^{(h)}}(t)$.  The homogeneous problem takes the form of a linear system with non-constant coefficients. Its solution may be formally written under general conditions in terms of a Magnus expansion (see e.g. \cite{Blanes09}). It then reads
\begin{equation}
\hat{\vec{W}}^{(h)}(t)= \ma{E}\mbox{xp} \left(  \ma{B}(t)\,
 \right) \cdot  \vec{\mathcal{C}}\,,
\end{equation}
 where 
 $\vec{\mathcal{C}} $ is a parameter to be varied, and $ \ma{E}\mbox{xp}  \left(  \ma{B}(T)
 \right)$ is the exponential of the matrix $\ma{B}$  defined as 
 \begin{equation}
 \ma{B}(t) = \ma{\Omega}_{\ma H}(t) - \left(\int_0^t K^2(\tau)\mbox{d}\tau \right) \ma{1} \:.
 \label{magnus}
\end{equation}
 In (\ref{magnus}), the matrix $\ma{\Omega}_{\ma H}$  is given in the form of a sum, namely
\begin{equation}
\ma{\Omega}_{\ma H}(t)= \sum_{k=1}^\infty \ma{\Omega}_{\ma H}^{(k)} (t)\:.
\label{hh}
\end{equation} 
The matrix $\ma{\Omega}_{\ma H}$ represents the Magnus expansion of ${\hat{\vec{W}}^{(h)}}$ and the $\ma{\Omega}_{\ma H}^{(k)}$ that appear in (\ref{hh}) are defined as
\begin{eqnarray}
\ma{\Omega}_{\ma H}^{(1)}  &=& \int_0^t \mbox{d} t_1\ma{ H}(t_1)\,, \nonumber\\
\ma{\Omega}_{\ma H}^{(2)} &=& \frac{1}{2} \int_0^t \mbox{d} t_1 \int_0^{t_1} \mbox{d}t_2\left[\ma{ H}(t_1),\ma{H}(t_2)\right]\,,\\
\ma{\Omega}_{\ma H}^{(3)}  &=& \frac{1}{6} \int_0^t \mbox{d} t_1 \int_0^{t_1} \mbox{d}t_2 \int_0^{t_2} \mbox{d} t_3 \left[\ma{ H}(t_1),\left[\ma{H}(t_2),\ma{H}(t_3)\right]\right] + \left[\ma{ H}(t_3),\left[\ma{H}(t_2),\ma{H}(t_1)\right]\right]\,,\nonumber\\
\ma{\Omega}_{\ma H}^{(4)}  &= &\ldots\,, \nonumber
\end{eqnarray}
 where the square brackets denote Lie brackets, so that for instance
$
\left[\ma{ H}(t_1),\ma{H}(t_2)\right] \equiv \ma{ H}(t_1) \cdot \ma{H}(t_2) - \ma{ H}(t_2) \cdot \ma{H}(t_1).
$
Applying the method of variation of parameters to this formal solution, and assuming that the disturbance velocity is zero at $t=0$, we are led to 
\begin{equation}
\vec{\hat{W}}(t) = \int_0^t K^2(\tau)  \:\ma{E}\mbox{xp} \left(  \ma{B}(t,\:\tau)
 \right)  \cdot \vec{\hat{T}}^{(0)}(\tau) \mbox{d}\tau\,,
\label{Sol}
\end{equation}
where 
\begin{equation}
 \ma{E}\mbox{xp} \left(  \ma{B}(t,\:\tau) \right) = \mbox{e}^{  - \int_\tau^t K^2(\tau') \text{d}\tau' }  \:\ma{E}\mbox{xp} \left(\ma{\Omega}_{\ma H}(t)\right) \cdot \ma{E}\mbox{xp} \left( -\ma{\Omega}_{\ma H}(\tau)\right)  \:.
\end{equation}
\\
Equation (\ref{Sol}) is the formal solution of the disturbance flow problem in Fourier space, expressed in the co-moving coordinate system.  However, to determine the inertial correction to the force acting on the body, we need to subtract the Stokeslet solution $\vec{\hat{T}}^{(0)}(t)$ from (\ref{Sol}). One way to achieve this is to perform an integration by parts of the latter. To this end, we first notice that 
\begin{equation}
K^2(\tau) \mbox{e}^{ - \int_\tau^t K^2(\tau') \text{d}\tau' }=  \frac{\mbox{d}}{\mbox{d}\tau}\left(\mbox{e}^{ - \int_\tau^t K^2(\tau') \text{d}\tau' }\right) \:.
\label{eq:B21}
\end{equation}
Using again the fact that the slip velocity is zero at $t=0$, we obtain after a few manipulations
\begin{equation}
\vec{\hat{W}}(t)   - \vec{\hat{T}}^{(0) }(t)   = 
 \int_0^t \ma{E}\mbox{xp} \left(  \ma{B}(t,\:\tau)
 \right)  \cdot  \left(\ma{H}(\tau)  \cdot \hat{\vec{T}}^{(0)}(\tau) -\frac{\mbox{d} \vec{\hat{T}}^{(0)}( \tau)  }{\mbox{d}\tau} \right)\mbox{d}\tau\:.
\label{eq:sol2}
\end{equation}
 The components of the Green tensor expressed in the co-moving coordinates at any time, $\tau$, may be written in the form
\begin{equation}
\hat{\ma{G}}_\mathcal{C}(\tau)  = \frac{1}{K^2(\tau)} \left( \ma{1}  + \frac{\ma{R}(\tau) \cdot \vec{K}^{\sf T}  \cdot \vec{K}}{K^2(\tau)}\right) \,.
\label{greent}
\end{equation}
\\
Then, making use of (\ref{eq:Stokeslet_curvilinear}) and (\ref{greent}), (\ref{eq:sol2})  may be re-cast as
\begin{equation}
\begin{split}
\vec{\hat{W}}(t)   - \vec{\hat{T}}^{(0) }(t)   =  
 \int_0^t \ma{E}\mbox{xp} \left(  \ma{B}(t,\:\tau)
 \right)  \cdot  \left(\ma{H}(\tau) \cdot \hat{\ma{G}}_\mathcal{C}(\tau) - \frac{\mbox{d} \hat{\ma{G}}_\mathcal{C}(\tau) }{\mbox{d}\tau} \right) \cdot \vec{F}^{(0)}(\tau) \mbox{d}\tau  \hspace{1cm}\\
  + \int_0^t \left(
  \hat{\ma{G}}_\mathcal{C}(t) \mbox{e}^{-\int_\tau^t K^2(\tau') \text{d}\tau'}  - \hat{\ma{G}}_\mathcal{C}(t) \mbox{e}^{-\int_\tau^t K^2(\tau') \text{d}\tau'}  
 - \ma{E}\mbox{xp} \left(  \ma{B}(t,\:\tau)\right)  \cdot  \hat{\ma{G}}_\mathcal{C}(\tau)  \right) \cdot  \frac{\mbox{d}\vec{F}^{(0)}(\tau)}{\mbox{d}\tau} \mbox{d}\tau\:.
\label{eq:sol3}
\end{split}
\end{equation}
In (\ref{eq:sol3}), the first two terms in the second integral cancel each other but they have been introduced artificially on purpose. Attention must be paid to the fact the Green-like matrix $\hat{\ma{G}}_\mathcal{C}(t)$ involved in these two terms is evaluated at the current time, $t$, instead of $\tau$. 
Integration again by parts, making use of (\ref{eq:B21}) and noting that
\begin{equation}
 \frac{\mbox{d}  }{\mbox{d}\tau}\left(\ma{E}\mbox{xp} \left(  \ma{B}(t,\:\tau)\right)\right) = \ma{E}\mbox{xp}\left(  \ma{B}(t,\:\tau)\right) \cdot \left( - \ma{H}(\tau) +K^2(\tau) \ma{1} \right)\,,
\end{equation}
it may be shown that 
\begin{eqnarray}
%\begin{split}
\nonumber
&&\int_0^t \left( \hat{\ma{G}}_\mathcal{C}(t) \mbox{e}^{-\int_\tau^t K^2(\tau') \text{d}\tau'}  
 - \ma{E}\mbox{xp} \left(  \ma{B}(t,\:\tau)\right)  \cdot  \hat{\ma{G}}_\mathcal{C}(\tau)  \right) \cdot  \frac{\text{d}\vec{F}^{(0)}(\tau)}{\mbox{d}\tau} \mbox{d}\tau   \\
 \label{interm}
&=&  - \int_0^t 
  K^2(\tau)\mbox{e}^{-\int_\tau^t K^2(\tau') \text{d}\tau'}  \hat{\ma{G}}_\mathcal{C}(t) \cdot  \vec{F}^{(0)}(\tau) \mbox{d}\tau  \\
  \nonumber
 &-& \int_0^t \left(
   \ma{E}\mbox{xp} \left(  \ma{B}(t,\:\tau)\right)  \cdot  \left(  \ma{H}(\tau) \cdot \hat{\ma{G}}_\mathcal{C}(\tau) - K^2(\tau) \hat{\ma{G}}_\mathcal{C}(\tau) - \frac{\mbox{d}\hat{\ma{G}}_\mathcal{C}(\tau)}{\mbox{d}\tau}\right)  \right)\cdot  \vec{F}^{(0)}(\tau) \mbox{d}\tau \:.
 %\end{split}
\end{eqnarray}
Thanks to (\ref{interm}), (\ref{eq:sol3}) may then be written in the final form 
\begin{equation}
\begin{split}
\vec{\hat{W}}(t)   - \vec{\hat{T}}^{(0) }(t)&    =   
 - \int_0^t   \hat{\ma{G}}_\mathcal{C}(t) \mbox{e}^{-\int_\tau^t K^2(\tau') \text{d}\tau'}  \cdot \frac{\mbox{d}\vec{F}^{(0)}(\tau)}{\mbox{d}\tau} \ \mbox{d}\tau \\
 & - \int_0^t K^2(\tau) \left( \mbox{e}^{-\int_\tau^t K^2(\tau') \text{d}\tau'} \hat{\ma{G}}_\mathcal{C}(t)   
 - \ma{E}\mbox{xp} \left(  \ma{B}(t,\:\tau)\right)  \cdot  \hat{\ma{G}}_\mathcal{C}(\tau)  \right) \cdot  \vec{F}^{(0)}(\tau) \mbox{d}\tau
\label{eq:sol4_1}
\end{split}
\end{equation}

\subsection{Solution in the Cartesian basis}
Integrating (\ref{eq:sol4_1}) over the three-dimensional $\vec{K}$-space yields the components of the disturbance force acting on the body in the $\vec{E}_i$-basis. However, the body slip velocity and acceleration are much more naturally expressed in the Cartesian $\vec{e}_i$-basis. This is why it is appropriate to re-write the disturbance solution in the Cartesian basis before integrating over the $\vec{k}$-space. 
To this end, we may use the fact that
\begin{equation}
\ma{F}(t)\cdot \ma{F}^{-1}(\tau) = \ma{F}(t-\tau)\:.
\end{equation}
In addition, we note that covariant components in the co-moving basis are linked to their counterpart in the Cartesian basis through the relation
\begin{equation}
\vec{K} = \vec{k} \cdot \ma{F}(t) \:,
\end{equation}
whereas contravariant components in the moving basis are linked to their counterpart in the Cartesian basis through 
\begin{equation}
\vec{\hat{W}}(t) - \vec{\hat{T}}^{(0)}(t) = \ma{F}^{-1}(t) \cdot ( \vec{\hat{w}}(t) - \vec{\hat{\mathcal{T}}}^{(0)}(t))\:.
\end{equation}
One may also notice that 
\begin{equation}
 \frac{\mbox{d} \vec{F}^{(0)}(\tau)}{\mbox{d}\tau} = \ma{F}^{-1}(\tau) \cdot \left(\frac{\mbox{d} \vec{f}^{(0)}(\tau)}{\mbox{d}\tau} - \ma{A} \cdot \vec{f}^{(0)}(\tau)\right)\:.
 \label{eq:B31}
\end{equation}
Here it is important to point out that (\ref{eq:B31}) is a component-to-component relation, not an intrinsic relation between tensors. %{\color{red}{Note also that  we have used here the fact that  $a^j_k = A^j_k$.}} 
 Finally, at the current time $t$, one has 
\begin{equation}
\ma{\hat{G}}_\mathcal{C}(t) = \ma{F}^{-1}(t) \cdot \ma{\hat{G}} \cdot \ma{F}(t)\:.
\end{equation}

Writing the solution of the problem in the Cartesian basis then leads to 
\begin{equation}
\begin{split} 
\vec{\hat{w}}(t) -\vec{\hat{\mathcal{T}}}^{(0)}(t)  = &   - \int_0^t \mbox{e}^{-\int_\tau^t K^2(t-\tau') \text{d}\tau'}  \ma{\hat{G}} \cdot \ma{F}(t-\tau)  \cdot \frac{\mbox{d} \vec{f}^{(0)}}{\mbox{d}\tau} \mbox{d}\tau  \\
& + \int_0^t  \mbox{e}^{-\int_\tau^t K^2(t-\tau') \text{d}\tau'} \ma{\hat{G}} \cdot \ma{F}(t-\tau) \cdot  \ma{A} \cdot \vec{f}^{(0)}  \mbox{d}\tau\\
&  - \int_0^t \mbox{e}^{-\int_\tau^t K^2(t-\tau') \text{d}\tau'}  K^2(t-\tau) \ma{\hat{G}} \cdot \ma{F}(t-\tau) \cdot \vec{f}^{(0)}  \mbox{d}\tau\\
&  + \int_0^t K^2(t-\tau) \ma{F}(t)\cdot \ma{E}\mbox{xp}\left(\ma{B}\right) \cdot \ma{G}_\mathcal{C}(\tau) \cdot  \ma{F}^{-1}(\tau) \cdot \vec{f}^{(0)}  \mbox{d}\tau\:,\\
\end{split}
\label{eq:sol4_2}
\end{equation}
\\
where we have used the fact that $K^2(\vec{K},\tau)=K^2(\vec{k},t-\tau)$ (see (\ref{eq:K2_lab})). Equation (\ref{eq:sol4_2}) may be further simplified by first writing the first integral on the right-hand side in the form
\begin{eqnarray}
\nonumber
%\begin{split}
 - \int_0^t \mbox{e}^{-\int_\tau^t K^2(t-\tau') \text{d}\tau'} & \ma{\hat{G}} \cdot \ma{F}(t-\tau) \cdot  \frac{\mbox{d} \vec{f}^{(0)}}{\mbox{d}\tau} \mbox{d}\tau
=   - \displaystyle\int_0^t \mbox{e}^{-\int_\tau^t K^2(t-\tau') \text{d}\tau'} \ma{\hat{G}}\cdot  \frac{\mbox{d} \vec{f}^{(0)}}{\mbox{d}\tau} \mbox{d}\tau \\
 +&\displaystyle\int_0^t \mbox{e}^{-\int_\tau^t K^2(t-\tau') \text{d}\tau'} \ma{\hat{G}} \cdot  \left(\ma{1} - \ma{F}(t-\tau)\right)  \cdot \frac{\mbox{d} \vec{f}^{(0)}}{\mbox{d}\tau} \mbox{d}\tau
%\end{split}
\label{splitsplit}
\end{eqnarray}
\\
The second integral on the right-hand side of (\ref{splitsplit}) can be integrated by parts as
\begin{eqnarray}
\nonumber
%\begin{split}
\int_0^t &\mbox{e}^{-\int_\tau^t K^2(t-\tau') \text{d}\tau'}  \ma{\hat{G}} \cdot  \left(\ma{1} - \ma{F}(t-\tau)\right)  \cdot \frac{\mbox{d} \vec{f}^{(0)}}{\mbox{d}\tau} \mbox{d}\tau  = - \displaystyle\int_0^t \mbox{e}^{-\int_\tau^t K^2(t-\tau') \text{d}\tau'}  \ma{\hat{G}} \cdot  \ma{F}(t-\tau) \cdot \ma{A} \cdot \vec{f}^{(0)}\mbox{d}\tau \\
&- \displaystyle\int_0^t \mbox{e}^{-\int_\tau^t K^2(t-\tau') \text{d}\tau'}  K^2(t-\tau) \ma{\hat{G}} \cdot  \left( \ma{1} - \ma{F}(t-\tau)\right) \cdot \vec{f}^{(0)}\mbox{d}\tau \:.
%\end{split}
\label{byparts}
\end{eqnarray}
\\
Inserting (\ref{splitsplit}) and (\ref{byparts}) in (\ref{eq:sol4_2}) then yields 
\begin{equation}
\begin{split} 
&\vec{\hat{w}}(t) -\vec{\hat{\mathcal{T}}}^{(0)}(t)   =    - \int_0^t \mbox{e}^{-\int_\tau^t K^2(t-\tau') \mbox{d}\tau'}  \ma{\hat{G}} \cdot  \frac{\mbox{d} \vec{f}^{(0)}}{\mbox{d}\tau} \mbox{d}\tau  \\
&  - \int_0^t \mbox{e}^{-\int_\tau^t K^2(t-\tau') \mbox{d}\tau'}  K^2(t-\tau) \left( \ma{\hat{G}} - \ma{F}(t) \cdot \ma{E}\mbox{xp}\left(\ma{\Omega}_{\ma{H}}(t,\tau)\right) \cdot \ma{G}_\mathcal{C}(\tau) \cdot \ma{F}^{-1}(\tau)\right) \cdot \vec{f}^{(0)}  \mbox{d}\tau\:.\\
\end{split}
\label{eq:sol4}
\end{equation}
We finally obtain (\ref{dw:fourier}) by expressing the last term within parentheses in the integrand of (\ref{eq:sol4}) thanks to the two relations
\begin{equation}
\ma{F}(\tau) \cdot \ma{G}_\mathcal{C}(\tau) \cdot \ma{F}^{-1}(\tau) = \frac{1}{K^2(t-\tau)} \left(\ma{1} 
- \frac{\ma{F}^{\sf T}(t-\tau) \cdot \vec{k} \vec{k} \cdot \ma{F}(t-\tau)}{K^2(t-\tau)}\right) \,,
\end{equation}
and 
\begin{equation}
\ma{F}(t) \cdot \ma{E}\mbox{xp}\left(\ma{\Omega}_\ma{H}(t,\:\tau)\right) \cdot   \ma{F}^{-1}(t)= 
\ma{E}\mbox{xp}\left(\ma{\Omega}_{[\ma{F}(t)\cdot \ma{H}(t,\:\tau) \cdot \ma{F}^{-1}(t)]}\right)  \equiv
\ma{Y}_2 \,.
\end{equation}

\clearpage
% \bibliographystyle{jfm}
% \bibliography{biblio}

\end{document}